\documentclass[twocolumn,twocolappendix,astrosymb]{aastex631}

\hypersetup{linkcolor=red,citecolor=blue,filecolor=cyan,urlcolor=magenta}
\def\lessim{\mathrel{\hbox{\rlap{\hbox{\lower4pt\hbox{$\sim$}}}\hbox{$<$}}}}
\def\grtsim{\mathrel{\hbox{\rlap{\hbox{\lower4pt\hbox{$\sim$}}}\hbox{$>$}}}}

\shorttitle{M83 Novae}
\shortauthors{Shafter et al.}

\graphicspath{{./}{figures/}}

\begin{document}

\title{A Survey of Novae in M83}

\correspondingauthor{Allen W. Shafter}
\email{ashafter@sdsu.edu}

\author[0000-0002-1276-1486]{A. W. Shafter}
\affiliation{Department of Astronomy and Mount Laguna Observatory,
San Diego State University, San Diego, CA 92182, USA}

\author[0000-0002-0835-225X]{K. Hornoch}
\affiliation{Astronomical Institute of the Czech Academy of Sciences, Fri\v{c}ova 298, CZ-251 65 Ond\v{r}ejov, Czech Republic}

\author[0000-0002-4319-8083]{J. Ben\'a\v{c}ek}
\affiliation{Center for Astronomy and Astrophysics, Technical University of Berlin, 10623 Berlin, Germany}

\author[0000-0002-5971-1136]{A. Gal\'ad}
\affiliation{Modra Observatory, Department of Astronomy, Physics of the Earth, and Meteorology, FMPI UK, Bratislava SK-84248, Slovakia}

\author[0000-0002-6384-0184]{J. Jan\'ik}
\affiliation{Department of Theoretical Physics and Astrophysics, Masaryk University, Kotl\'a\v{r}sk\'a 2, 611 37 Brno, Czech Republic}

\author[0000-0002-3130-4168]{J. Jury\v{s}ek}
\affiliation{Institute of Physics of the Czech Academy of Sciences, Prague, Czech Republic}
\affiliation{D\'epartement d'Astronomie, Universit\'e de Geneve, Chemin d'Ecogia 16, CH-1290 Versoix, Switzerland}

\author{L. Kotkov\'a}
\affiliation{Astronomical Institute of the Czech Academy of Sciences, Fri\v{c}ova 298, CZ-251 65 Ond\v{r}ejov, Czech Republic}

\author[0000-0003-0311-3222]{P. Kurf\"{u}rst}
\affiliation{Department of Theoretical Physics and Astrophysics, Masaryk University, Kotl\'a\v{r}sk\'a 2, 611 37 Brno, Czech Republic}

\author[0000-0002-1330-1318]{H. Ku\v{c}\'akov\'a}
\affiliation{Astronomical Institute of the Czech Academy of Sciences, Fri\v{c}ova 298, CZ-251 65 Ond\v{r}ejov, Czech Republic}
\affiliation{Research Centre for Theoretical Physics and Astrophysics, Institute of Physics, Silesian University in Opava, Bezru\v{c}ovo n\'am. 13, CZ-74601 Opava, Czech Republic}

\author[0000-0001-6098-6893]{P. Ku\v{s}nir\'ak}
\affiliation{Astronomical Institute of the Czech Academy of Sciences, Fri\v{c}ova 298, CZ-251 65 Ond\v{r}ejov, Czech Republic}

\author[0000-0001-7080-1565]{J. Li\v{s}ka}
\affiliation{Central European Institute of Technology - Brno University of Technology (CEITEC BUT), Purky\v{n}ova 656/123, CZ-612 00, Brno, Czech Republic}
\affiliation{Variable Star and Exoplanet Section of the Czech Astronomical Society, V\'ide\v{n}sk\'a 1056, CZ-142 00, Praha-Libu\v{s}}

\author[0000-0002-3304-5200]{E. Paunzen}
\affiliation{Department of Theoretical Physics and Astrophysics, Masaryk University, Kotl\'a\v{r}sk\'a 2, 611 37 Brno, Czech Republic}

\author[0000-0002-7602-0046]{M. Skarka}
\affiliation{Astronomical Institute of the Czech Academy of Sciences, Fri\v{c}ova 298, CZ-251 65 Ond\v{r}ejov, Czech Republic}
\affiliation{Department of Theoretical Physics and Astrophysics, Masaryk University, Kotl\'a\v{r}sk\'a 2, 611 37 Brno, Czech Republic}

\author[0000-0002-7434-9518]{P. \v{S}koda}
\affiliation{Astronomical Institute of the Czech Academy of Sciences, Fri\v{c}ova 298, CZ-251 65 Ond\v{r}ejov, Czech Republic}

\author[0000-0002-4387-6358]{M. Wolf}
\affiliation{Astronomical Institute, Charles University, Prague, V Hole\v{s}ovi\v{c}k\'ach 2, CZ-180 00 Praha 8, Czech Republic}

\author[0000-0001-9383-7704]{P. Zasche}
\affiliation{Astronomical Institute, Charles University, Prague, V Hole\v{s}ovi\v{c}k\'ach 2, CZ-180 00 Praha 8, Czech Republic}

\author[0000-0001-6231-3350]{M. Zejda}
\affiliation{Department of Theoretical Physics and Astrophysics, Masaryk University, Kotl\'a\v{r}sk\'a 2, 611 37 Brno, Czech Republic}

\begin{abstract}
The results of the first synoptic survey of novae in the barred spiral
and starburst galaxy, M83 (NGC~5236), are presented.
A total of 19 novae and one background supernova were discovered
during the course of a nearly seven-year survey comprised of over 200
individual nights of observation between 2012 December 12 and 2019 March 14.
After correcting for the limiting magnitude and the spatial and
temporal coverage of the survey, the nova rate in M83 was found to
be $R=19^{+5}_{-3}$~yr$^{-1}$. This rate,
when normalized to the $K$-band luminosity of the galaxy, yields a
luminosity-specific nova rate,
$\nu_K = 3.0^{+0.9}_{-0.6}\times10^{-10}$~yr$^{-1}~L_{\odot,K}^{-1}$.
The spatial distribution of the novae is found to be
more extended than the overall
galaxy light suggesting that the observed
novae are likely dominated by a disk population.
This result is consistent with the observed nova light curves
which reveal that the M83 novae are on average
more luminous at maximum light and fade faster when compared with novae
observed in M31.
Generally, the more luminous M83
novae were observed to fade more rapidly, with the complete sample
being broadly consistent with
a linear Maximum-Magnitude vs Rate of Decline relation.
\end{abstract}

\keywords{Cataclysmic variable stars (203) -- Classical Novae (251) -- Galaxies (573) -- Novae (1127) -- Time Domain Astronomy (2109)}

\section{Introduction}  \label{sec:intro}

Nova eruptions are the result of quasi-periodic
thermonuclear runaways (TNRs) on the surfaces
of accreting white dwarfs in semidetached binary systems
\citep[e.g., see][and references therein]{2016PASP..128e1001S},
with eruptions recurring on timescales as short as a year
\citep{2014ApJ...793..136K}\footnote{
Novae where more than one eruption has been recorded (i.e.,
systems with recurrence times less than of order a century)
are collectively referred to a ``Recurrent Novae", although
the terminology is somewhat misleading given that all systems
are believed to be recurrent.}.
Novae are among the most luminous optical transients known, with
absolute magnitudes at the peak of the eruption averaging M$_V\sim-7.5$, and
reaching M$_V\sim-10$ for the most luminous systems. As a result, they
can be seen to great distances and have been studied in external
galaxies for more than a century
\citep[e.g., in M31,][]{1917PASP...29..210R,1929ApJ....69..103H}.

The observed properties of novae are predicted theoretically to
depend strongly on the structure of the progenitor binary system.
The mass of the white dwarf and the rate of accretion
onto its surface are the most important parameters, ultimately
determining the ignition mass required to initiate the TNR
\citep[e.g.,][]{1982ApJ...253..798N,2005ApJ...628..395T,2014ApJ...793..136K}.
Systems with high mass white
dwarfs accreting at high rates require the lowest ignition masses,
and thus have the shortest recurrence times between
successive eruptions. The small ignition masses result in eruptions
that eject relatively little mass
resulting in a rapid photometric evolution (i.e., they produce
``fast" novae).

The mass accretion rate
is strongly influenced by the evolutionary
state of the companion star, with evolved
stars typically transferring mass
to the white dwarf at a higher rate compared with systems containing
main-sequence companions. The amplitudes of the eruptions are
also strongly dependent on the nature of the companion star, being
as small as $\sim$5 magnitudes as in the case of the
M31 recurrent nova M31N~2008-12a \citep{2017ApJ...849...96D},
or as large as $\sim$20 mag
as was observed for the Galactic nova, V1500 Cyg \citep{1987SSRv...45....1D}.
Finally, it is also thought that
the chemical composition of the accreted material may also play
an important role in determining the observed properties of nova eruptions
\citep[e.g.,][and references therein]{2016PASP..128e1001S}.

Given that the nature of the nova eruptions are predicted to depend
sensitively on the properties of the progenitor binary, it is reasonable
to expect that the observed properties of a novae in a given galaxy might vary 
with the underlying stellar population. In particular, the specific
{\it rate\/} of nova eruptions can be expected to be much higher in a 
population of novae containing higher mass white dwarfs, where the
average recurrence times are relatively short.
An early attempt to explore
this question was undertaken by \citet{1997ApJ...481..127Y} who computed
population synthesis models that predicted that young stellar populations,
which contain on average more massive white dwarfs,
should produce nova eruptions at a higher rate compared with older populations.
Thus, late-type, low mass galaxies, with a recent history of
active star formation were predicted to be more prolific nova producers
compared with older, quiescent galaxies.

To date, nova rates have been estimated in well over a dozen
galaxies \citep[e.g., see][and references therein]
{2014ASPC..490...77S,2019enhp.book.....S,2020A&ARv..28....3D}.
Taken together, the results do not
suggest a simple relationship
between a galaxy's specific nova rate (usually taken to be its $K$-band
luminosity-specific rate, $\nu_K$) and its
dominant stellar population (as reflected by its integrated $B-K$ color).
Early work by
\citet{1990AJ.....99.1079C,2000ApJ...530..193S,2004ApJ...612..867W}
failed to find any correlation between $\nu_K$
and Hubble type; however, \citet{1994A&A...287..403D} argued that the bluer,
late-type systems such as the Magellanic Clouds and M33, had higher
specific nova rates compared with earlier type galaxies.
More recently, \citet{2016ApJS..227....1S}
and \citet{2017RNAAS...1...11S}
have analyzed archival {\it HST\/} imaging data of M87
and made a compelling case
that the specific nova rate in this giant elliptical galaxy is at least
as high, and perhaps higher, than that found in spiral galaxies and the LMC.
Given the
uncertainties inherent in measuring extragalactic nova rates, particularly
in galaxies other than the Magellanic Clouds (where the rates are relatively
well constrained by the Optical Gravitational
Lensing Experiment \citep[OGLE,][]{2016ApJS..222....9M}, it's
fair to say that the question of whether the specific
rates vary systematically with
the underlying stellar population has yet to be answered definitively.

In an attempt to shed further light on how the underlying
stellar population may affect observed nova properties, we have undertaken a
multi-year survey of novae in the grand-design spiral
M83 (NGC~5236) -- also known as the Southern Pinwheel galaxy --
a metal-rich, late-type barred spiral galaxy of
morphological type SAB(s)c \citep{1991rc3..book.....D}.

\begin{deluxetable}{lccc}
\tablecolumns{4}
\tablecaption{Log of Observations\label{tab:log}}
\tablehead{\colhead{UT Date} & \colhead{Julian Date} & \colhead{Limiting mag} & \\
\colhead{(yr~mon~day)} & \colhead{(2,450,000+)} & \colhead{($R$)} & \colhead{Notes\tablenotemark{a}}
}
\startdata
2012 12 12.361  & 6273.861  & 22.3 & 1 \cr
2012 12 18.368  & 6279.868  & 22.7 & 1 \cr
2012 12 22.352  & 6283.852  & 22.5 & 1 \cr
2012 12 23.345  & 6284.845  & 22.9 & 1 \cr
2012 12 28.314  & 6289.814  & 21.8 & 1 \cr
\enddata
%\tablenotetext{a}{This Table is a stub.}
\tablecomments{Table~\ref{tab:log} is published in its entirety in the machine-readable format. A portion is shown here for guidance regarding its form and content.}
\tablenotetext{a}{All observations were made with the 1.54-m Danish Telescope at La Silla. Observer(s): (1) K. Hornoch
}
\end{deluxetable}

\begin{figure}
\includegraphics[angle=0,scale=0.40]{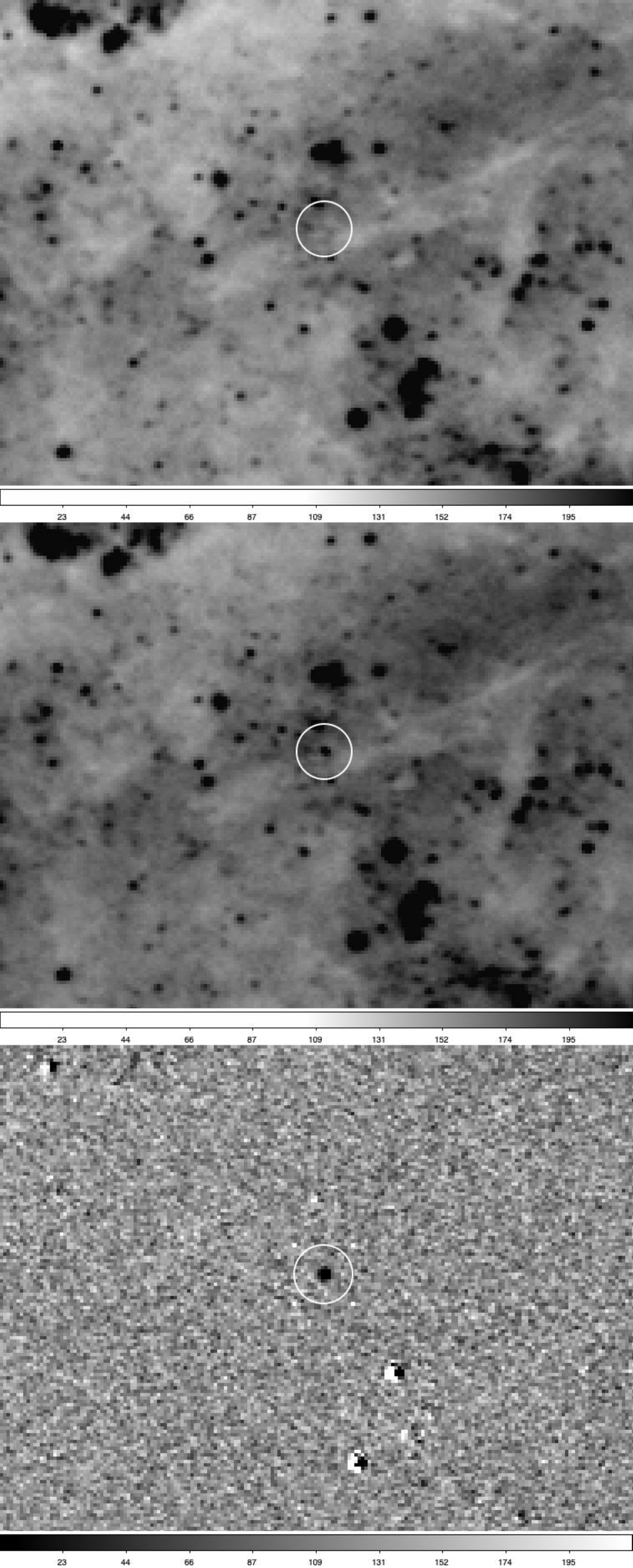}
\caption{Top panel: A $75''\times57''$ portion of a median-combined
image from our survey taken on the night of 2014 Feb 11 UT.
Middle panel: An image of the same region of the galaxy one
month later on 2014 Mar 11 UT. Bottom panel: The difference
of the two images as computed by the ISIS image subtraction
software, clearly showing the nova M83N 2014-03a.
}
\label{fig:novdet}
\end{figure}

\section{Observations}  \label{sec:obs}

Our survey for novae in M83 spanned approximately seven years between
2012 December 12 and 2019 March 14. During this time we acquired
a total of 205 nightly images of the galaxy.
All observations
were acquired with the 1.54-m Danish reflector at the La Silla observatory
using the DFOSC $2048\times2048$ CCD imager.
The telescope-detector combination resulted in
final images
covering a square area approximately $13.5'$ on a side,
with a spatial resolution of 0.396 arcsec~pixel$^{-1}$.
In visual light, M83
has an apparent size of $12.9'\times11.5'$, assuring that our
observations cover essentially all of the galaxy. However, as
described later in section~\ref{ssec:lsnr}, the outer halo
of M83 may extend slightly beyond our survey's spatial coverage.

The vast majority of
our images were taken through a broad-band $R$ filter, with occasional
exposure taken through an $I$ filter. We chose to conduct
our primary survey in the $R$-band both
because novae develop strong H$\alpha$ emission shortly after eruption,
which adds to the flux in the $R$-band, and because the quantum efficiency
of the CCD detector reaches a peak near the $R$-band.
A complete log of our observations is
given in Table~\ref{tab:log}.

\subsection{Image Processing}

All images were pipeline processed in the usual manner
by first subtracting the bias and dark current,
and then flat-fielding the individual images to remove
the high-frequency, pixel-to-pixel variations
using APHOT \citep[a synthetic aperture photometry and astrometry
software developed by M. Velen and P. Pravec at
the Ond\v{r}ejov observatory,][]{1994ExA.....5..375P}.
To eliminate cosmic-ray artifacts in our nightly images,
we obtained a series of 120-s images that were
later spatially registered and median stacked using
SIPS\footnote{\tt https://www.gxccd.com/cat?id=146\&lang=409}
to produce a final image for a given night of observation
intended for nova searching.
Photometric and astrometric measurements of the novae were done using
APHOT on spatially
registered and stacked nightly images.

\begin{figure}
\plotone{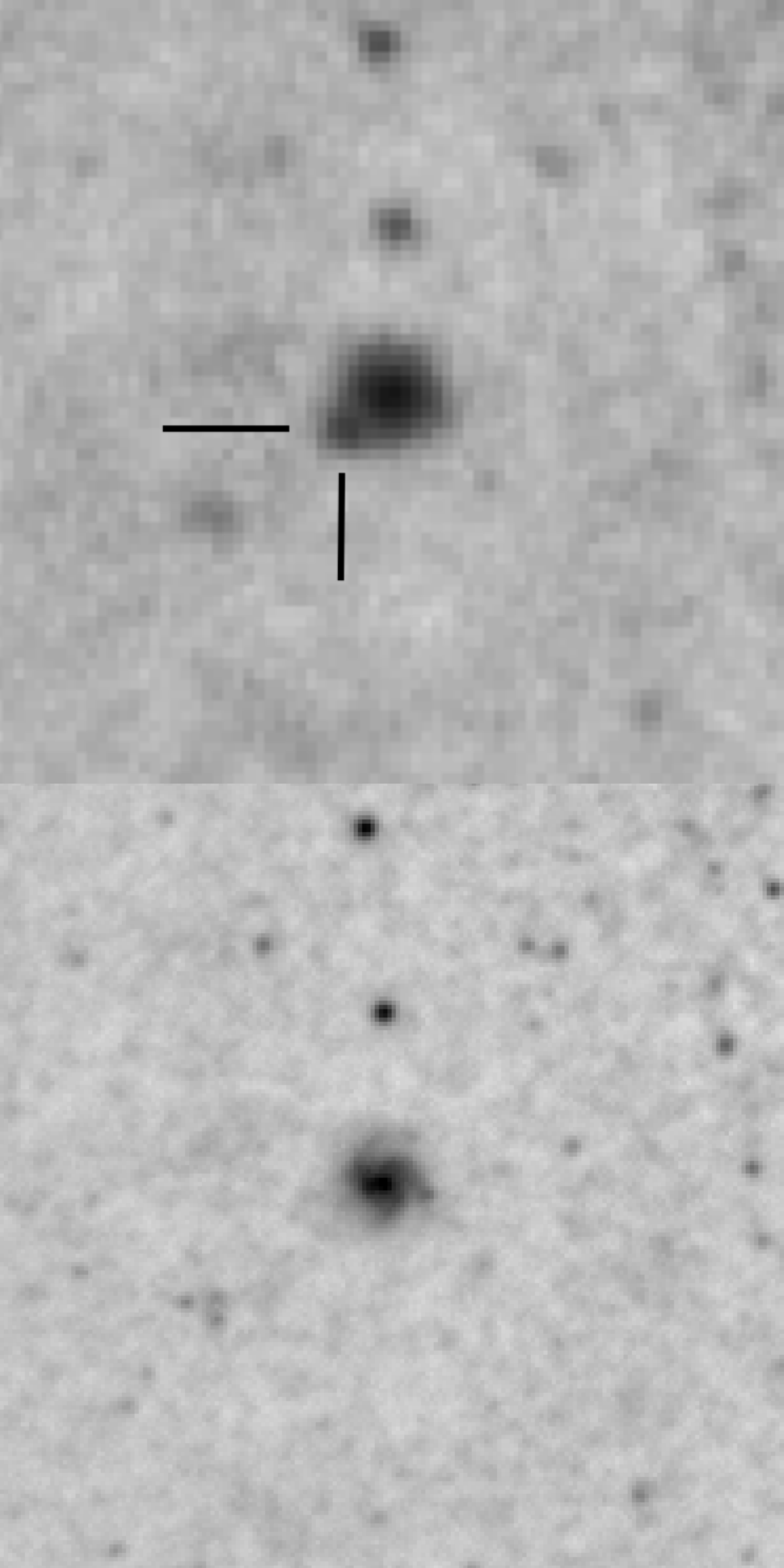}
\caption{Top panel: A $30''\times30''$ portion of our survey
image from the night of 2013 Jan 18 UT (North up, East left).
The likely supernova
is located approximately $2''$ E and $1.6''$ S of the center of
the anonymous spiral galaxy. For comparison, the bottom panel shows an identical
$30''\times30''$ portion of deep Sloan $r$-band image of the galaxy field
taken on 2010 Sep 28 UT by J. L. Prieto
using the MEGACAM on the 6.5-m Magellan II - Clay telescope
at Las Campanas Observatory.
}
\label{fig:posssn}
\end{figure}

\begin{figure}
\includegraphics[angle=0,scale=0.33]{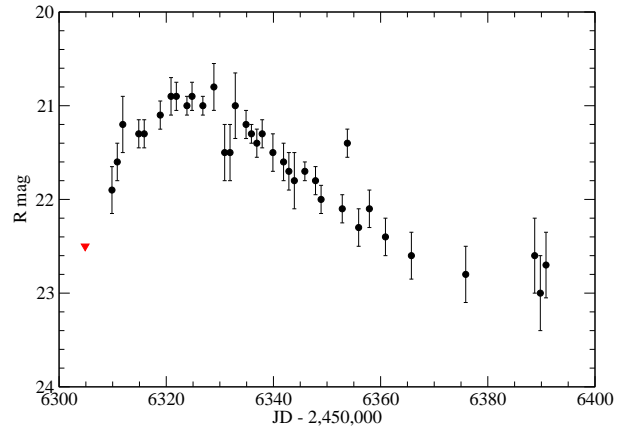}
\caption{The light curve of the transient source near the
anonymous background galaxy. The proximity of the transient source
to the galaxy, coupled with the slow rise to maximum light ($\grtsim 10$~d),
strongly suggests that the object is a supernova (likely of Type~Ia)
in the background
galaxy, and not a nova in M83. The red triangle represents an upper limit
on the flux when the transient was not detected.
}
\label{fig:snlc}
\end{figure}

\begin{deluxetable*}{ccccrrc}
\tablecolumns{7}
\tablecaption{M83 Novae \label{tab:novae}}
\tablehead{& & \colhead{R.A. (2000.0)} & \colhead{Decl. (2000.0)} & \colhead{R.A. offset} & \colhead{Decl. offset} & \\
\colhead{Nova \#} & \colhead{Nova name} & \colhead{(h~~m~~s)} & \colhead{($^{\circ}$~~$'$~~$''$)} & \colhead{$\Delta\alpha$~cos($\delta$) ($''$)} & \colhead{$\Delta\delta$ ($''$)} & \colhead{Notes\tablenotemark{a}}
}
\startdata
1& M83N 2013-01a & 13 36 44.30 & $-29~50~33.9$ &$ -216.2$ &$   82.8$ &  1 \cr 
2& M83N 2013-01b & 13 37 03.42 & $-29~53~39.8$ &$   32.5$ &$ -103.1$ &  2 \cr
3& M83N 2013-01c & 13 37 06.75 & $-29~56~15.5$ &$   75.8$ &$ -258.8$ &  3 \cr
4& M83N 2013-01d & 13 37 10.51 & $-29~45~19.6$ &$  124.8$ &$  397.1$ &  3 \cr
5& M83N 2013-03a & 13 36 58.76 & $-29~53~22.6$ &$  -28.1$ &$  -85.9$ &  4 \cr
6& M83N 2013-04a & 13 37 05.80 & $-29~48~11.4$ &$   63.5$ &$  225.3$ &  5 \cr
7& M83N 2014-01a & 13 37 17.86 & $-29~53~47.7$ &$  220.3$ &$ -111.0$ &  6 \cr
8& M83N 2014-01b & 13 37 05.81 & $-29~56~02.0$ &$   63.6$ &$ -245.3$ &  3 \cr
9& M83N 2014-01c & 13 37 15.48 & $-29~56~10.9$ &$  189.3$ &$ -254.2$ &  3 \cr
10& M83N 2014-03a & 13 37 08.88 & $-29~50~16.6$ &$  103.6$ &$  100.1$ &  7 \cr
11& M83N 2015-01a & 13 36 35.15 & $-29~56~41.0$ &$ -335.1$ &$ -284.3$ &  8 \cr
12& M83N 2015-04a & 13 37 08.42 & $-29~51~13.5$ &$   97.6$ &$   43.2$ &  9 \cr
13& M83N 2016-02a & 13 36 53.01 & $-29~56~19.9$ &$ -102.8$ &$ -263.2$ &  10 \cr
14& M83N 2016-02b & 13 36 45.52 & $-29~55~57.2$ &$ -200.2$ &$ -240.5$ &  10 \cr
15& M83N 2016-03a & 13 36 36.15 & $-29~51~44.9$ &$ -322.2$ &$   11.8$ &  11 \cr
16& M83N 2018-01a & 13 37 09.78 & $-29~56~57.6$ &$  115.2$ &$ -300.9$ &  12 \cr
17& M83N 2018-02a & 13 37 01.26 & $-29~55~48.2$ &$    4.4$ &$ -231.5$ &  13 \cr
18& M83N 2019-02a & 13 37 03.82 & $-29~48~21.7$ &$   37.7$ &$  215.0$ &  14 \cr
19& M83N 2019-03a & 13 37 18.18 & $-29~48~29.0$ &$  224.6$ &$  207.7$ &  15 \cr
\enddata
\tablenotetext{a}{
(1) \citep{2013ATel.4723....1H};
(2) \citep{2013ATel.4732....1H}, \citep[see also][for spectroscopic classification]{2013ATel.4734....1P};
(3) K. Hornoch;
(4) K. Hornoch and M. Skarka;
(5) A. W. Shafter, K. Hornoch, and M. Wolf;
(6) K. Hornoch and J. Vra\v{s}til;
(7) K. Hornoch and M. Zejda;
(8) K. Hornoch and E. Paunzen;
(9) K. Hornoch, E. Paunzen, and M. Zejda;
(10) K. Hornoch, P. Zasche, E. Paunzen, and J. Li\v{s}ka;
(11) K. Hornoch, M. Wolf, L. Pilar\v{c}\'ik, and J. Vra\v{s}til;
(12) \citep{2018ATel11240....1H};
(13) \citep{2018ATel11443....1H};
(14) \citep{2019ATel12539....1H};
(15) \citep{2019ATel12564....1H}.}
\end{deluxetable*}

\section{Nova Detection} \label{sec:novdet}

Novae are transient sources that can be effectively identified
through careful comparisons between images
from synoptic imaging surveys having a variable cadence
such as the one we have conducted.
Coarse temporal coverage is
sufficient to identify transient sources, but sufficiently dense
coverage is required to measure the light curves. The light curves
provide assurance that the detected transients are indeed novae and
not some other variable objects, such as Luminous Blue Variables (LBVs), or
the more common Luminous Red Variables (i.e., Mira variables)
that can mimic novae in poorly sampled surveys.
Novae can also
be distinguished from other variable stars through their $V-I$ color,
which is the reason we augmented our data with occasional $I$-band images.
The strong H$\alpha$ emission contributing to the $R$-band flux
results in a $V-I$ color that
is significantly bluer than that of the extremely red
Mira variables, which are
characterized by $V-I \grtsim 2$ \citep[e.g., see][]{2019ApJ...884...20B}.

Nova candidates were identified using two different procedures: First,
through a direct comparison (blinking) of images from different epochs,
and secondly, through a comparison of
images from differing epochs
using the image subtraction software, ISIS \citep{1998ApJ...503..325A}.
Direct comparison of the images proved to be a very
effective technique for identifying novae in the outer regions
of the galaxy; however, in regions of the galaxy with high surface brightness
(i.e., within $\sim3'$ of the nucleus and in some very dense regions of the 
spiral arms) we had to modify our approach.
To detect novae in regions of high background,
we first created smoothed (median-filtered) images by
sliding a $11\times11$ pixel box across
the image, pixel-by-pixel, replacing the central
pixel by the median of all 121 pixels in the box.
This median-smoothed image was then subtracted from the original image to
produce a median-subtracted image with greatly reduced background
variations.
These median-subtracted images were once again blinked by eye.

In addition to blinking median-subtracted images, we also employed
image subtraction routines from the ISIS image processing package,
which increased our sensitivity to
novae in regions of high background.
Figure~\ref{fig:novdet}
shows an example of our ISIS nova detection procedure.
The top
panel shows a $75''\times57''$ portion of a median-combined image
from 2014 Feb 11 UT.
The middle panel shows an image of the same region of the galaxy one
month later on 2014 Mar 11 UT. The nova, M83N 2014-03a, is
visible in the March image, and clearly visible in the ISIS subtracted image
at the bottom.

We discovered a total of 19 novae
over the course of our seven-year survey of M83. Their positions and
offsets from the center of M83 (R.A. = $13^h 37^m 00.^s919$, Decl. = $-29^\circ 51' 56.''74$, J2000)
are given in Table~\ref{tab:novae}. These detections represent the first
novae to be reported in M83.
It is worth noting that during our inspection of the images
we found one transient source
located extremely close ($\sim2''$~E and $\sim1.6''$~S) to the center of an
anonymous background spiral galaxy (see Figure~\ref{fig:posssn}).
The background galaxy, which is located at R.A. = $13^h~37^m~19.59^s$,
Decl. = $-29^{\circ}~53'~47.1''$, lies
$242.3''$~E and $110.3''$~S from the center of M83.
While it is possible that this transient could be a nova in M83, given its
proximity to the nucleus of the spiral, we consider it far
more likely to have been a supernova in the galaxy discovered
serendipitously during the course of our survey \citep{2013ATel.4747....1H}.
This assessment
is backed up by the light curve of the transient (see Figure~\ref{fig:snlc}),
which shows the relatively slow rise to peak brightness characteristic
of supernovae, but not novae. A comparison with the supernova light curve
templates given in \citet{1985AJ.....90.2303D} suggests that the transient
is likely a supernova of Type~Ia. In view of these considerations,
we have excluded this source from our final tally of M83 novae.

\subsection{Photometry} \label{ssec:phot}

\begin{deluxetable}{lcccc}
\tablecolumns{5}
\tablecaption{M83 Nova Photometry \label{tab:phot}}
\tablehead{\colhead{(UT Date)} & \colhead{(JD - 2,450,000)} & \colhead{Mag} & \colhead{Unc.} & \colhead{Band}
}
\startdata
\cutinhead{2013-nova1 = 2013-01a}
2012 06 01.138 &  6079.638 & [22.6  &     &R   \cr
2012 12 12.361 &  6273.861 & [22.3  &     &R   \cr
2012 12 18.368 &  6279.868 & [22.7  &     &R   \cr
2012 12 23.345 &  6284.845 & [22.9  &     &R   \cr
2012 12 28.314 &  6289.814 &  21.0  &0.15 &R   \cr
2012 12 28.316 &  6289.816 &  21.1  &0.2  &I   \cr
\enddata
\tablecomments{Table~\ref{tab:phot} is published in its entirety in the machine-readable format. A portion is shown here for guidance regarding its form and content.}
\end{deluxetable}

\begin{figure*}[ht!]
\includegraphics[angle=0,scale=1.0]{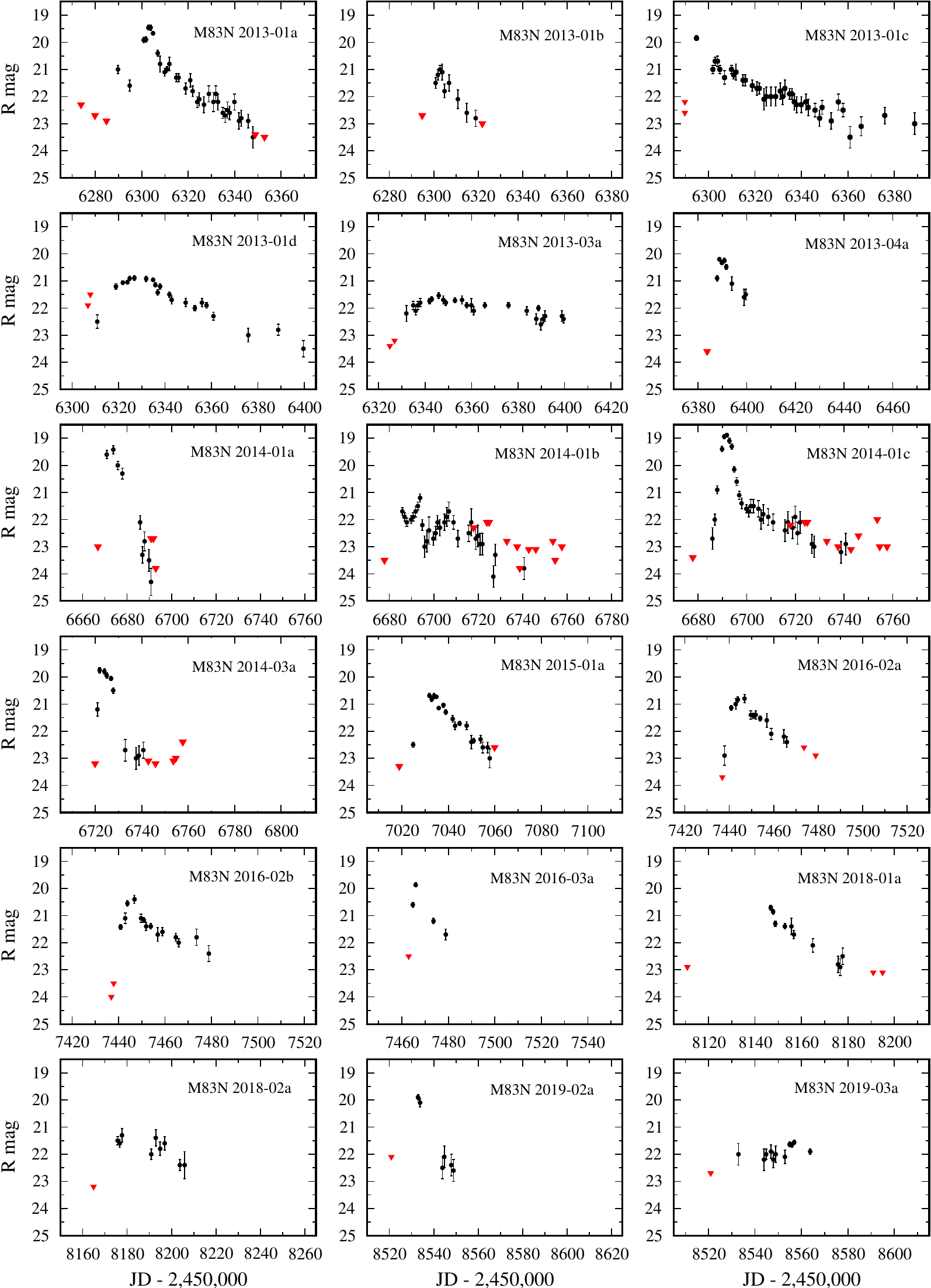}
\caption{Light curves for 18 of the 19 novae discovered in our survey
plotted with consistent axes scales in order to better reveal the
relative photometric properties of the novae. The one nova not shown
is M83N 2015-04a, where the nova was only seen in one epoch
($R=21.4\pm0.25$ on JD 2,457,120.915).
The red triangles represent upper limits
on the flux at times when the novae were not detected.}
\label{fig:lightcurves}
\end{figure*}

\begin{figure*}
\includegraphics[angle=0,scale=0.11]{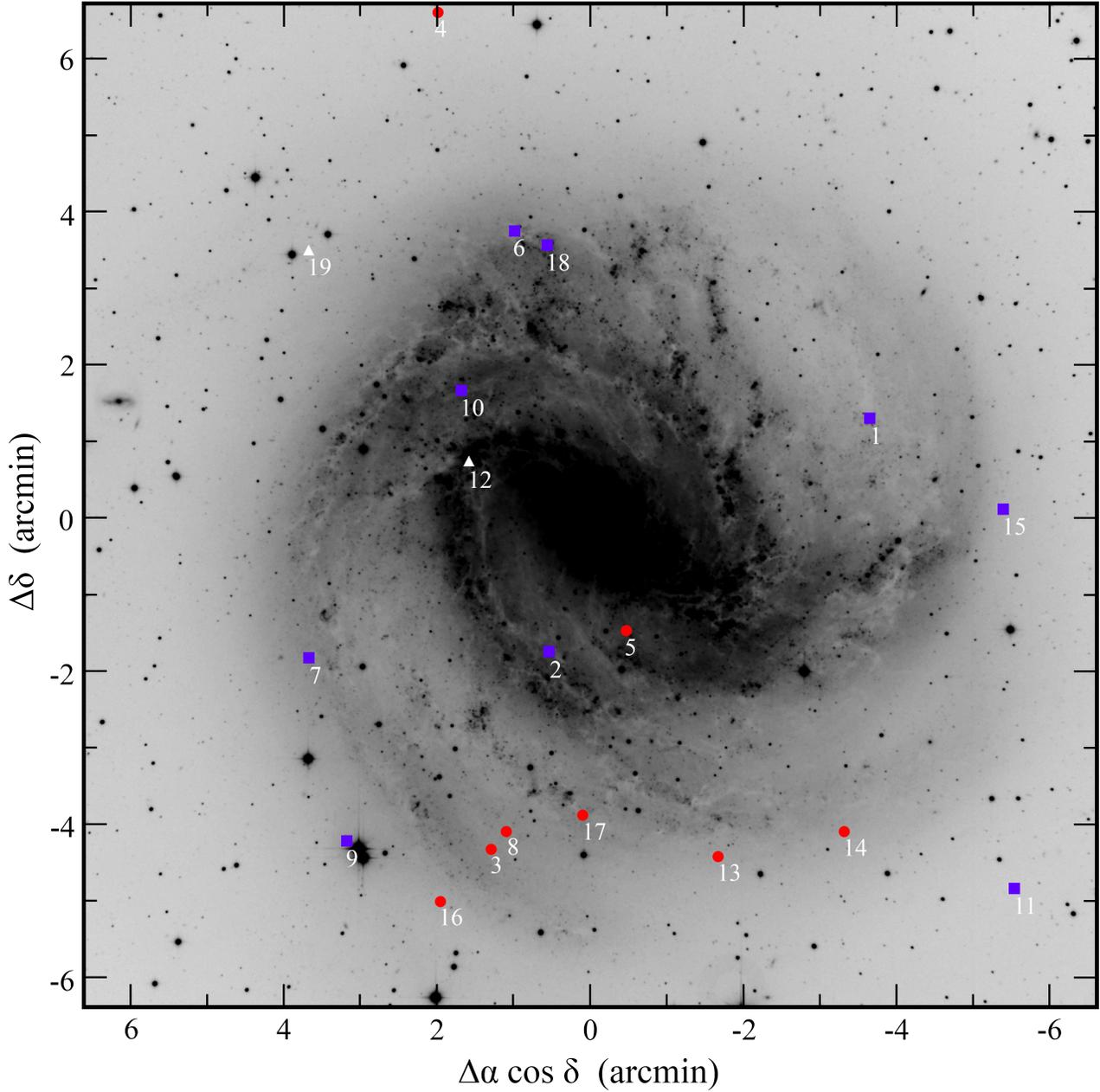}
\caption{The positions of the 19 novae discovered in our survey superimposed
on an image of M83. The blue squares represent relatively ``fast" novae
with $t_2<25$~d, while the red circles show ``slow" novae with $t_2>25$~d.
The white triangles show the two novae without sufficient light curve
coverage to determine a rate of decline from peak brightness.
We find no correlation between speed class and spatial position
within the galaxy, and it is clear that the novae are more spatially
extended than is the background galaxy light.}
\label{fig:spatdist}
\end{figure*}

Instrumental magnitudes for all nova candidates were determined
by summing the fluxes in 2$''$-in diameter circular apertures
in all epochs in which they were visible.
Calibrated $R$-band magnitudes
were then determined by differential photometry
with respect to a set of six secondary standard stars in the M83 field.
Both the $R$-band and $I$-band magnitudes of the secondary standard stars were
photometrically calibrated by us using the same instrumentation as we used
for the survey. As primary standards we used the stars and their magnitudes
published in \citet{1992AJ....104..340L}.
We observed both the primary standards and the M83
field during one night under excellent photometric conditions to get properly
calibrated the secondary standards in the field of M83 which allow us to obtain
photometry of the novae also in non-photometric nights.

The calibrated magnitudes of all 19 novae discovered as part of our survey,
on all nights where they
were visible, are presented in Table~\ref{tab:phot}.
The temporal sampling of our survey was sufficient to
produce useful $R$-band light curves for all but one of the 19 novae.
The light curves for these 18 novae having multiple epochs of observation
are shown in Figure~\ref{fig:lightcurves}. The properties of the
light curves (i.e., their peak magnitudes and fade rates)
will be explored further in section~\ref{sec:lcproperties}.

\section{The Spatial Distribution of M83 Novae} \label{sec:spatdist}

\begin{deluxetable}{ccccc}
\tabletypesize{\scriptsize}
\tablewidth{0pt}
\tablecolumns{5}
\tablecaption{M83 Surface Photometry\tablenotemark{\scriptsize a}\label{tab:surfphot}}
\tablehead{\colhead{Semimajor axis} & \colhead{$\Sigma_R$} & & \colhead{Semimajor axis} & \colhead{$\Sigma_R$} \\
 \colhead{(arcsec)}  & \colhead{(mag/arcsec$^2$)} & & \colhead{(arcsec)} & \colhead{(mag/arcsec$^2$)}
}
\startdata
   0.4 & 15.90 && 104.1 & 20.29 \cr
   2.7 & 16.01 && 107.7 & 20.25 \cr
   4.4 & 16.42 && 110.7 & 20.25 \cr
   7.1 & 16.98 && 113.8 & 20.25 \cr
   9.3 & 17.59 && 117.4 & 20.25 \cr
  11.5 & 17.91 && 121.8 & 20.29 \cr
  13.7 & 18.17 && 125.3 & 20.33 \cr
  16.3 & 18.45 && 128.8 & 20.38 \cr
  18.1 & 18.65 && 131.5 & 20.46 \cr
  20.7 & 18.80 && 137.7 & 20.55 \cr
  22.1 & 18.91 && 142.5 & 20.61 \cr
  24.7 & 18.99 && 148.2 & 20.68 \cr
  25.6 & 19.06 && 157.1 & 20.74 \cr
  27.4 & 19.19 && 165.0 & 20.85 \cr
  29.6 & 19.25 && 175.6 & 20.94 \cr
  31.8 & 19.30 && 185.7 & 21.02 \cr
  33.5 & 19.38 && 199.4 & 21.18 \cr
  36.2 & 19.45 && 209.1 & 21.26 \cr
  38.4 & 19.51 && 222.4 & 21.37 \cr
  41.0 & 19.60 && 236.0 & 21.50 \cr
  43.2 & 19.64 && 245.3 & 21.61 \cr
  46.8 & 19.68 && 250.2 & 21.72 \cr
  49.9 & 19.73 && 259.0 & 21.74 \cr
  52.9 & 19.81 && 268.2 & 21.85 \cr
  56.9 & 19.86 && 277.1 & 21.95 \cr
  60.9 & 19.92 && 287.7 & 22.11 \cr
  65.3 & 19.99 && 299.1 & 22.28 \cr
  69.7 & 20.03 && 320.0  & 22.56\tablenotemark{\scriptsize b} \cr
  73.2 & 20.10 && 350.0  & 22.96\tablenotemark{\scriptsize b} \cr
  78.5 & 20.14 && 400.0  & 23.63\tablenotemark{\scriptsize b} \cr
  86.0 & 20.20 && 450.0  & 24.31\tablenotemark{\scriptsize b} \cr
  90.4 & 20.25 && 500.0  & 24.98\tablenotemark{\scriptsize b} \cr
  93.5 & 20.27 && \dots  & \dots \cr
  98.4 & 20.27 && \dots  & \dots \cr
\enddata
\tablenotetext{\scriptsize a}{D\scriptsize igitized from \citet{2000ApJS..131..441K}. P.A. = 80$^{\circ}$; Ellipticity, $b/a = 0.9$ assumed for all values.}
\tablenotetext{\scriptsize b}{\scriptsize extrapolated value}
\end{deluxetable}

\begin{figure}
\includegraphics[angle=0,scale=0.32]{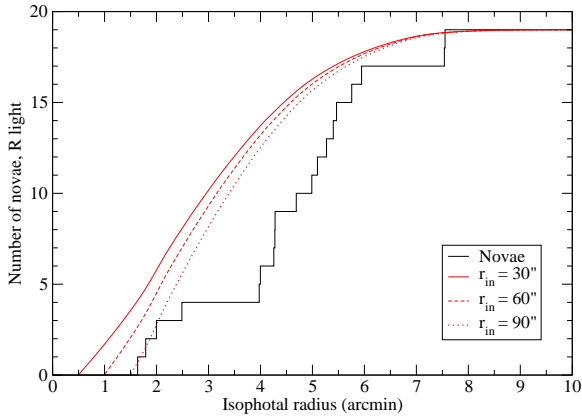}
\caption{The cumulative distribution of the isophotal radii of the
19 novae (black histogram) compared with the cumulative galaxy
$R$-band light for three representative values of the
inner radius, $R_{in}$ (shown in red).}
\label{fig:raddis}
\end{figure}

Figure~\ref{fig:spatdist} shows the spatial distribution
of the 19 novae discovered in M83
superimposed on a
(negative) image of the
galaxy.
It is immediately apparent that the novae seem to be more
spatially extended than the galaxy light. We can explore this impression more
quantitatively by comparing the cumulative distribution of the novae
with that of the background $R$-band light,
as shown in Figure~\ref{fig:raddis}.
The cumulative background light
has been determined
by integrating the surface photometry from Table~\ref{tab:surfphot},
which we have derived
from digitizing the radial surface brightness profiles given in Figure~4
of \citet{2000ApJS..131..441K}.
Since the high surface brightness near the nucleus of the galaxy renders us
effectively blind to novae within $\sim30''$ of the center of
the galaxy, and likely significantly incomplete within $60''$,
we have considered three cumulative light distributions starting with
inner radii at $R_{\mathrm in}=30''$, $R_{\mathrm in}=60''$,
and $R_{\mathrm in}=90''$.
Regardless of the adopted inner radius, the cumulative
background light clearly falls off faster than the observed nova distribution.
This result is formally confirmed by Kolmogorov-Smirnov (K-S)
tests ($p=1.6\times10^{-3}$, $p=3.7\times10^{-3}$, and $p=1.1\times10^{-2}$
for cases where we considered inner radii of
$R_{in}=30''$, $R_{in}=60''$, and $R_{in}=90''$, respectively), suggesting
we can reject the null hypothesis (i.e., the nova and light distributions
were drawn from the same parent distribution) with $\grtsim99\%$ confidence.
Thus, it appears that the novae detected in M83 are primarily associated
with a more extended disk population of the galaxy.

Before considering possible explanations for why the novae
do not appear to follow the overall background light in M83, it is important
to rule out the possibility that we may be missing a significant
fraction of novae in the inner regions
of the galaxy due to the difficulty in detecting them against
the high central surface brightness. We explore
this possibility below in our determination of the nova
rate in M83, which also depends critically on
the overall completeness of our survey.

\section{The Nova Rate in M83} \label{sec:novrat}

As transient objects, novae
are visible for a limited time that depends both on the intrinsic properties
of the novae themselves (e.g., their peak luminosities, fade rates, and
spatial location within the galaxy) and on
parameters inherent to the survey itself; specifically, the temporal
sampling and survey depth (the effective limiting magnitude). 
Whether or not a given nova in M83 can potentially be detected
depends on a combination of its
peak apparent magnitude,
its position within the galaxy, and on the limiting magnitude
of our survey images at that location in the galaxy.
Then, whether it will actually be detected
depends on the time it erupted relative
to the dates of our observations and the rate of decline in the nova's
brightness. Given a population of novae with a variety of light curve
properties (peak luminosities and fade rates), distributed at different
positions within the galaxy, and observed at different times, the only
practical way of determining the number of novae we can expect to see
in our survey given an intrinsic rate, $R$, is to conduct numerical
(Monte Carlo) simulations. Before the simulations can be performed,
the first step is to determine the effective completeness of the survey
at a given magnitude, $C(m)$.

\subsection{The Effective Limiting Magnitude of the Survey}

A determination of the limiting magnitude of our survey images
is complicated by the fact that the galaxy background surface brightness
is highly
variable and that the spatial distribution of the novae
cannot be assumed to be uniform across the survey images.
Thus, no one limiting magnitude can represent the coverage of a given image.
We have approached this problem following the procedure described
in our earlier work on NGC~2403 \citep{2012ApJ...760...13F}. Specifically,
we have conducted artificial nova tests
on a representative image (hereafter the fiducial image)
under the assumption that the spatial
distribution of the artificial novae follows the background light of the galaxy.

The artificial novae
were generated using tasks in the IRAF DAOPHOT package, which enabled us to
match the point-spread-functions of the real stars in the image.
Using the routine \texttt{addstar}, the fiducial
image was then seeded with 100 artificial novae having
apparent magnitudes randomly distributed within each of a total
of eight, 0.5-mag wide, bins.
For each of the magnitude bins, the artificial novae were distributed
randomly, but with a spatial density that was constrained
to follow the integrated background galaxy light of M83.
We then searched for the artificial novae using the same procedures
that we employed in identifying the real novae. The completeness
at the fainter magnitudes was somewhat higher using the ISIS
image subtraction analysis, but was generally consistent with the
results from a direct comparison of median-subtracted images.

\begin{figure}
\includegraphics[angle=0,scale=0.33]{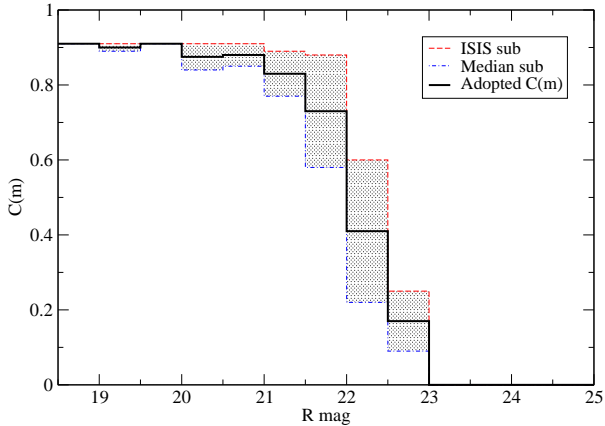}
\caption{The survey completeness as a function of the $R$-band
apparent magnitude, $C(m)$, as determined from our artificial
nova tests. The red dashed line shows the completeness
from the ISIS image subtraction procedure, the blue dot-dash
line shows the completeness using the median-subtraction technique,
with the heavy black line shows the average completeness
in each magnitude bin.
The shaded region shows the range of completeness in each magnitude bin
bounded by the two search techniques and
used in our Monte Carlo nova rate simulations.}
\label{fig:com}
\end{figure}

The fraction of novae recovered from the two search techniques
in each magnitude bin yielded the completeness functions
shown in Figure~\ref{fig:com}.
Given that we discovered the same number of novae in M83 employing
both techniques,
we have chosen to take the average completeness in a given magnitude bin
(the heavy line in Figure~\ref{fig:com})
to form the basis of the completeness function, $C(m)$.
To allow for uncertainty in magnitude bins where the completeness functions
differ (the shaded regions),
we randomly sample allowed values of $C(m)$ in the analysis to follow.
This completeness function can be generalized
to any epoch, $i$, of observation
by applying a shift,
$\Delta m_i$ ($= m_{lim,0} - m_{lim,i}$), which represents the difference in
the limiting magnitudes (as measured by the faintest star that could
be reliably detected near the perimeter of the image away from the galaxy)
of our fiducial image and that of the $i$-th epoch image.
Thus, for any epoch, $i$, we have $C_i(m) = C(m+\Delta m_i)$.

\subsection{The Monte Carlo Simulation}

As in our earlier extragalactic nova
studies \citep[e.g.,][]{2012ApJ...760...13F,2010ApJ...720.1155G,2008ApJ...686.1261C,2004ApJ...612..867W}, we have
employed a Monte Carlo simulation
to compute the number of M83 novae
that we would expect to
observe during the course of our survey.

For a given assumed annual nova rate, $R$, we begin by producing trial novae
erupting at random times throughout the time span covered by our survey,
each having a peak luminosity and fade rate that has been selected at random
from a large sample of known $R$-band light curve parameters.
Ideally, we would like to use light curve parameters specific to the
full population of M83 novae, but such an unbiased sample does not exist.
Instead, we have used the M83 light curve parameters from our observations
given in Table~\ref{tab:lcparam}, augmented with additional $R$-band light curve
parameters from the M31 light curves observed by \citet{2011ApJ...734...12S}.

\begin{deluxetable}{lcccr}
\tablecolumns{5}
\tablecaption{Light Curve Parameters\label{tab:lcparam}}
\tablehead{\colhead{Nova} & \colhead{m$_R$} & \colhead{M$_R$\tablenotemark{a}} & \colhead{$f_R$} & \colhead{$t_2(R)$} \\
\colhead{(M83N)} & \colhead{(peak)} & \colhead{(peak)} & \colhead{(mag~d$^{-1}$)} & \colhead{(d)}
}
\startdata
2013-01a&$19.58\pm0.04$&$-8.90 \pm0.08$&$0.103\pm0.004$&$ 19.3\pm0.7$ \cr
2013-01b&$21.11\pm0.12$&$-7.37 \pm0.14$&$0.115\pm0.018$&$ 17.5\pm2.7$ \cr
2013-01c&$20.27\pm0.06$&$-8.21 \pm0.09$&$0.046\pm0.002$&$ 43.2\pm2.1$ \cr
2013-01d&$20.80\pm0.04$&$-7.69 \pm0.08$&$0.036\pm0.002$&$ 54.8\pm2.9$ \cr
2013-03a&$21.66\pm0.05$&$-6.82 \pm0.08$&$0.013\pm0.002$&$157.5\pm21.1$ \cr
2013-04a&$20.16\pm0.05$&$-8.32 \pm0.08$&$0.123\pm0.015$&$ 16.3\pm2.1$ \cr
2014-01a&$19.41\pm0.11$&$-9.08 \pm0.13$&$0.256\pm0.014$&$  7.8\pm0.4$ \cr
2014-01b&$21.68\pm0.06$&$-6.80 \pm0.09$&$0.033\pm0.004$&$ 60.6\pm7.0$ \cr
2014-01c&$18.69\pm0.05$&$-9.80 \pm0.08$&$0.326\pm0.012$&$  6.1\pm0.2$ \cr
2014-03a&$19.48\pm0.06$&$-9.00 \pm0.09$&$0.169\pm0.012$&$ 11.9\pm0.9$ \cr
2015-01a&$20.63\pm0.03$&$-7.85 \pm0.08$&$0.085\pm0.004$&$ 23.5\pm1.0$ \cr
2016-02a&$20.83\pm0.07$&$-7.65 \pm0.10$&$0.072\pm0.008$&$ 27.7\pm2.9$ \cr
2016-02b&$20.68\pm0.06$&$-7.80 \pm0.09$&$0.058\pm0.005$&$ 34.7\pm3.1$ \cr
2016-03a&$19.90\pm0.08$&$-8.59 \pm0.11$&$0.157\pm0.013$&$ 12.8\pm1.1$ \cr
2018-01a&$20.89\pm0.05$&$-7.60 \pm0.09$&$0.067\pm0.006$&$ 29.9\pm2.5$ \cr
2018-02a&$21.47\pm0.10$&$-7.01 \pm0.12$&$0.024\pm0.007$&$ 83.3\pm22.6$ \cr
2019-02a&$19.91\pm0.09$&$-8.57 \pm0.11$&$0.181\pm0.016$&$ 11.0\pm1.0$ \cr
\enddata
\tablenotetext{a}{Assuming a distance modulus
$\mu_{\circ}(\mathrm{M}83) = 28.34\pm0.14$
\citep{2016AJ....152...50T} and a
foreground $R$-band extinction of 0.14 mag \citep{2011ApJ...737..103S}.}
\end{deluxetable}

Adopting a distance modulus to M83 of
$\mu_{0} = 28.34 \pm 0.14$, which represents
the mean of the Cepheid and the tip of the
red giant branch distances from the recent study by \citet{2016AJ....152...50T},
enables us to compute
the expected apparent magnitude distribution at any given epoch, $i$,
during our survey, $n_i(m,R)$. To account for uncertainty in the
distance, our numerical simulations also randomly select values
of the distance modulus normally distributed about the mean value.
The number of novae expected to be detectable during the course of
our survey, $N_{obs}(R)$, can then be
computed by convolving the simulated apparent magnitude distribution
with the completeness function, $C_i(m)$, and then summing
over all epochs of observation:

\begin{equation}
N_{\mathrm obs}(R) = \sum_{i}\sum_{m}{C_i(m)~n_i(m,R)}.
\label{eqn:NovaNum}
\end{equation}

\begin{figure}
\includegraphics[angle=0,scale=0.33]{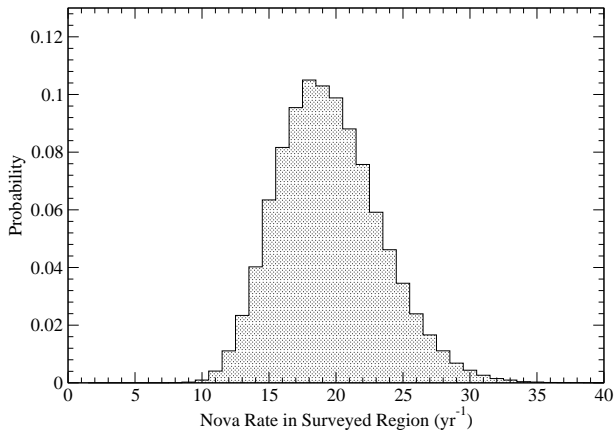}
\caption{The probability distribution for the nova rate in M83
as determined from our Monte Carlo simulations. The peak of the
distribution represents the most likely
nova rate in M83. Assuming a bi-Gaussian form for the probability
distribution yields a nova rate and associated $1\sigma$ uncertainty
of $R_{M83} = 18^{+5}_{-3}$ novae per year.}
\label{fig:montecarlo}
\end{figure}

%\noindent
The intrinsic nova rate in M83, $R$, and its uncertainty
can now be determined through a comparison of the
number of novae found in our survey, $n_{\mathrm obs}=19$,
with the number of novae predicted by equation (1).
We explored trial nova rates ranging from $R=1$ to $R=50$ novae per year,
repeating the numerical simulation 10$^{5}$ times for each trial
value of $R$. The number
of matches, $M(R)$, between the predicted number of observable novae,
$N_{obs}(R)$, and the actual number of novae discovered in our survey,
$n_{obs}$ = 19, was recorded for each trial value of $R$.
The number of matches
was then normalized by the total number of matches for all $R$
to give the probability distribution function, $P(R)=M(R)/\sum_{R}{M(R)}$
shown in
Figure~\ref{fig:montecarlo}. The most probable nova rate in the
portion of the galaxy covered by our survey images
is 18$^{+5}_{-3}$~yr$^{-1}$, where the error
range (1$\sigma$) for the asymmetrical probability distribution
has been computed assuming it
can be approximated by a bi-Gaussian function.

The $K$-band photometry of M83 from the
Two-Micron All Sky Survey (2MASS) suggests that the
extended halo may extend out to a distance of $r_{tot}\sim8.5'$
from the center of M83. Thus, it
is possible that our survey images may be missing a small fraction
($\lessim$5\%) of the light from the extended halo of M83. Under the
assumption that we are sampling $0.95\pm0.05$ of the total M83 light,
we estimate the global nova rate for M83 to be
$R_{M83} = 19^{+5}_{-3}$~yr$^{-1}$.

\subsection{The Luminosity-Specific Nova Rate} \label{ssec:lsnr}

In order to compare the nova rates between different galaxies or
different stellar populations, the rates must first be suitably normalized.
Ideally, it would be appropriate to normalize the rates
by the mass of stars in the region surveyed, but the mass cannot be
measured directly. As a proxy for the mass in stars, it has become
standard practice to normalize the nova rate by the infrared $K$-band
luminosity of the galaxy. In the case of M83, the integrated apparent
$K$-band magnitude
as measured by 2MASS is
$K=4.62\pm0.03$. Given the distance modulus, $\mu_{0} = 28.34 \pm 0.14$,
and taking the absolute $K$-band magnitude of the sun to be $M_K=3.27\pm0.02$
\citep{2018ApJS..236...47W},
we find that M83 has an
absolute magnitude in the $K$-band of $M_K=-23.72$, and a corresponding
$K$-band luminosity of
$(6.32\pm0.84)\times10^{10}$~L$_{\odot,K}$. Since we estimate that our survey
covers $0.95\pm0.05$ of the entire galaxy where we have found
an overall nova rate
of $18^{+5}_{-3}$ yr$^{-1}$, we arrive at a $K$-band luminosity-specific
nova rate for M83 of
$\nu_K = (3.0^{+0.9}_{-0.6})\times10^{-10}$~yr$^{-1}$~L$_{K,\odot}^{-1}$.

\begin{figure}
\includegraphics[angle=0,scale=0.32]{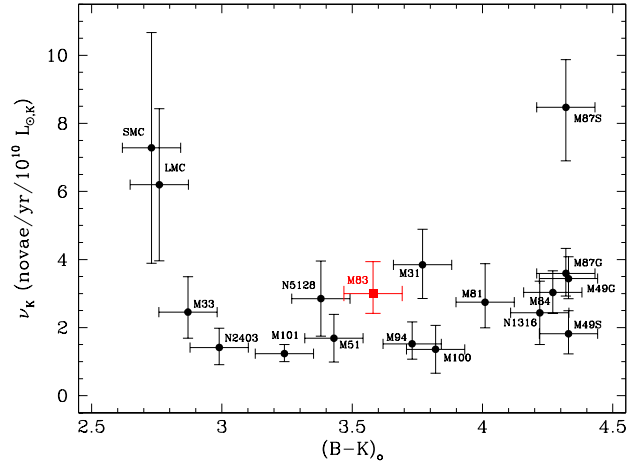}
\caption{The luminosity-specific nova rates of all 16 galaxies
where nova rates are available plotted as a function 
of the galaxy $B-K$ color. The data have been taken from Table~7.3
of \citet{2019enhp.book.....S} with the exception of the
value for M83 from the current study, which is shown as a red square.}
\label{fig:lsnr}
\end{figure}

As recently reviewed by \citet{2019enhp.book.....S} and
\citet{2020A&ARv..28....3D} prior to the present study, luminosity-specific
nova rates had been measured for a total of 15 external galaxies.
Figure~\ref{fig:lsnr} shows our value of
$\nu_K$ for M83, along with 
the values for the other 15 galaxies
taken from Table~7.3 in \citet{2019enhp.book.....S},
plotted as a function of the $B-K$ color of the host galaxy.
The specific nova rate for M83 is consistent with those
of other spiral galaxies with measured nova rates.
As we have
noted in previous studies, despite the relatively high $\nu_K$
values reported for
the Magellanic Clouds and M87 -- galaxies with
very different Hubble types -- there is no compelling
evidence that $\nu_K$ varies systematically with the
underlying stellar population.

\section{Light Curve Properties}  \label{sec:lcproperties}

As described in section~\ref{ssec:phot},
the temporal coverage during the course of our survey was sufficiently
dense to enable us to measure $R$-band
light curves for 18 of the 19 M83 novae that were detected.
These light curves, drawn from an equidistant sample of novae,
offer a rare opportunity to
explore the relationship between
a nova's peak luminosity and its rate of decline from maximum light,
and thus test whether or not the novae
obey the canonical (but recently questioned) Maximum-Magnitude
versus Rate-of-Decline (MMRD) relation.

\subsection{The MMRD Relation} \label{ssec:mmrd}

The MMRD relation for novae was first introduced by \citet{1945PASP...57...69M},
who discovered that the most luminous members of
a sample of 30 (mostly Galactic) novae
faded more quickly than did their fainter counterparts\footnote{In his
1945 paper McLaughlin referred to the MMRD relation as the ``life-luminosity"
relation for novae.}.
He was able to quantify a linear MMRD relation of the form,
$M = a+b~\mathrm{log}(t_3)$, where $a$ and $b$ are fitting parameters
and $t_3$ is the time it takes a nova to fade 3 magnitudes from
maximum light (in recent years $t_2$, which is more easily measured,
is often used as an alternative).
Over the years the MMRD relation has been calibrated
many times, both in the Galaxy \citep[e.g.,][]{1985ApJ...292...90C,
2000AJ....120.2007D},
and in nearby galaxies such as M31
\citep[e.g.,][]{1989AJ.....97.1622C,2011ApJ...734...12S},
and has often been used
as a means for determining the distances to novae where the
apparent magnitude at maximum light and the fade rate have been measured.

Over the years it has become increasingly apparent that
there is significant scatter in the MMRD, with the existence of the
relation itself being called into question, initially by
\citet[e.g.,][]{2011ApJ...735...94K} who found that a number of M31
novae observed with the Palomar Transient Factory (PTF) appeared to fall
below the canonical MMRD relation
(but, see \citet{2020A&ARv..28....3D} for a different interpretation.).
Recently, it has become clear that a small subset of novae, typically those
with massive white dwarfs that are accreting at high rates
(e.g., recurrent novae)
fade rapidly despite having relatively low peak luminosities,
and thus deviate sharply from the MMRD relation. A good
example is the M31 recurrent nova M31N 2008-12a \citep{2014A&A...563L...9D,2014ApJ...786...61T}.
Given that such systems do in fact exist,
it is clear that not all novae will follow a universal MMRD relation.
On this basis it has been recently suggested that
the notion of an MMRD relation should be abandoned altogether
\cite{2018MNRAS.481.3033S}. However, the question of
whether it is useful to continue to refer to an MMRD relation
would seem to depend on the relative frequency of outliers.
If the so-called ``Faint and Fast" (FFN) novae, such as
M31N 2008-12a, are intrinsically rare, then continuing to refer
to an MMRD might make sense when considering the behavior of
the majority of (non recurrent) novae.
On the other hand, if such systems are
relatively common, but just missed in most surveys that
lack the depth and cadence to discover them, then perhaps the existence of
an MMRD relation would be best considered as resulting from an
observational selection effect. In either case, it appears that
broadly speaking, luminous novae do on average fade more quickly than
do their low luminosity counterparts.

\begin{figure}
\includegraphics[angle=0,scale=0.33]{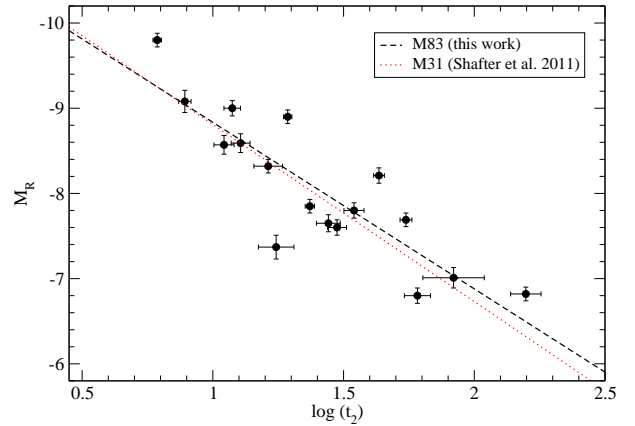}
\caption{The peak absolute $R$-band magnitudes for the 17 M83 novae
with complete light curves are plotted as a function of the log
of the fade rate, as measured by the time in days
a nova takes to fade by 2 magnitudes in the $R$-band from its peak
luminosity, $t_2(R)$. The M83 nova sample clearly displays a correlation
between peak nova luminosity and $\mathrm{log}(t_2)$, as shown by the
black dashed line.
The $R$-band MMRD relation for M31 from \citet{2011ApJ...734...12S}
(red dotted line)
is shown for comparison. The two relations are remarkably similar.}
\label{fig:mmrd}
\end{figure}

Figure~\ref{fig:mmrd} shows the MMRD relation for the 17 novae in M83 where
the light curves were sufficiently complete to allow a measurement
of the peak magnitude and the rate of decline (here measured
as $t_2$, the time for the nova to fade 2 mags from peak).
Although there is significant scatter as expected, there is no
question that a general trend, where the more luminous novae
fade more quickly, is apparent. The best-fitting linear MMRD relation
is given by: $\mathrm{M}_R = (-10.79\pm0.42) + (1.96\pm0.29)~\mathrm{log}~t_2$.
The M31 $R$-band MMRD relation
from the study of \citet{2011ApJ...734...12S} is shown for
comparison, and is remarkably similar\footnote{MMRD relations are sometimes
fit using a more complicated {\tt arctan} function, which has been shown
to provide a somewhat better fit to data for M31 and the LMC
\citep[e.g., see][]{1995ApJ...452..704D}.
Given the limited data in the present study,
we have chosen to employ the traditional linear fit.}.
Whether or not we are missing
a putative population of FFN novae in M83 is unknown. The question
can only be answered by future studies having greater depth and cadence,
such as those that will be possible with the Large Synoptic Survey
Telescope \citep[LSST,][]{2019ApJ...873..111I}.

\subsection{Comparison with the M31 nova population}

As part of a comprehensive spectroscopic and photometric survey
of novae in M31, \citet{2011ApJ...734...12S}
determined $R$-band light curve parameters
for a total of 42 novae and found that, as in the case of
M83, the M31 nova sample also generally followed a MMRD relation,
albeit with significant scatter \citep[see Table~6 and
Figure~19 in][]{2011ApJ...734...12S}.
The M31 novae are characterized
by a mean absolute magnitude, $\langle M_R \rangle =-7.55\pm0.14$ and
$\langle t_2(R) \rangle =38^{+6}_{-5}$~d,
while their M83 counterparts from Table~\ref{tab:lcparam} yield
$\langle M_R \rangle =-8.06\pm0.21$
and $\langle t_2(R) \rangle =25^{+6}_{-5}$~d\footnote{Since the
$t_2(R)$ distributions are highly asymmetric, the values
of $\langle t_2(R) \rangle$ are computed from the average
of the log$~t_2(R)$ values for each galaxy.}.

\begin{figure}
\includegraphics[angle=0,scale=0.33]{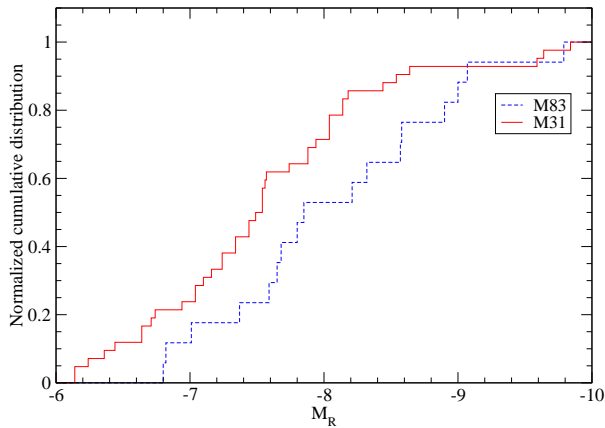}
\caption{The cumulative distribution of the $R$-band
absolute magnitudes of M83 novae compared with their M31
counterparts. A K-S test reveals that the two distributions
differ with 94\% confidence.
}
\label{fig:cumr}
\end{figure}

\begin{figure}
\includegraphics[angle=0,scale=0.33]{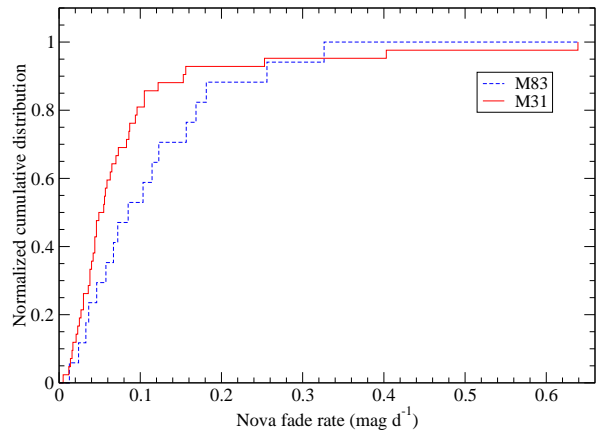}
\caption{The cumulative distribution of the fade rate
of M83 novae in the $R$-band compared with their M31
counterparts. A K-S test reveals that the two distributions
differ, but with just 74\% confidence.
}
\label{fig:cumf}
\end{figure}

It is interesting to compare the cumulative distributions of
the peak luminosities and fade rates from each galaxy, as shown in
Figures~\ref{fig:cumr} and~\ref{fig:cumf}.
The results of K-S tests
show that the peak luminosity and fade rate distributions
differ between each galaxy with 94\% and 74\% confidence, respectively.
Thus, there is some evidence that the novae in the M83 sample are, on average,
more luminous at maximum light and perhaps fade somewhat more rapidly
compared with the M31 sample. A possible explanation for this difference
is that M31 is an earlier Hubble type galaxy, SA(s)b, and the nova
population observed by \citet{2011ApJ...734...12S} was predominately a
bulge population. On the other hand, as discussed earlier,
our M83 sample appears to be primarily associated with the galaxy's disk.
Taken together, figures~\ref{fig:cumr} and~\ref{fig:cumf}
provide additional support
for the existence of two populations of novae, as originally
suggested for Galactic systems three decades ago by
\citet{1990LNP...369...34D} and
\citet{1998ApJ...506..818D} based on photometric and spectroscopic
observations, respectively.

\section{Discussion}  \label{sec:disc}

\subsection{Uncertainty in the Derived Nova Rate}

The determination of extragalactic nova rates is challenging,
with many sources of uncertainty that must be properly considered
in the analysis. Among these, uncertainty in the determination of the
survey completeness is perhaps the most important. As described earlier,
the completeness -- the fraction of novae erupting in the galaxy
that can be detected in a given survey -- depends quite sensitively on the
intrinsic properties of the novae and their actual distribution
within the galaxy under investigation. Since neither of these factors are
known {\it a priori}, assumptions concerning both must always be made.

To guard against potential bias in our M83 nova sample,
we chose to include
light curve parameters from previous observations of novae in M31
in our Monte Carlo nova rate simulations. As a check on the sensitivity
of the final nova rate to the inclusion of the M31 light curve parameters,
we have also performed the analysis using only the light curve parameters
from the M83 novae discovered in our survey. Despite the differences
in the light curve properties between the two samples discussed in the
previous section,
the final nova rate determination is essentially unchanged
(the rate ranges from 17~yr$^{-1}$ to 20~yr$^{-1}$ when
the analysis is restricted to the M83 and M31 light curve parameters,
respectively). 
In this case, the insensitivity of the nova rate
to the choice of light curve parameters
results from the fact that the M83 novae generally
follow an MMRD relation. The M83 novae are on average
brighter compared with their M31 counterparts,
and thus easier to detect in our simulations, but they fade
generally more quickly, so they are observable on average for less time.
Thus, the two effects tend to cancel, leaving the final nova rate
insensitive to the choice of light curve parameters.

The form of the nova spatial distribution that we have adopted in the artificial
nova completeness experiments can also affect the final
nova rate computation.
In the absence of strong
evidence to the contrary, the default approach -- and the one we have followed
here -- is to assume that
the spatial density of novae follows the surface brightness of the
galaxy (i.e., more light, more novae).
However, in the case of M83,
it appears that this assumption may be violated.
As shown in Figure~\ref{fig:spatdist} the {\it observed\/}
M83 nova spatial distribution
is significantly more extended than the galaxy's $R$-band light.
If this result accurately reflects the
intrinsic nova spatial distribution (i.e., we are not missing novae
in the inner regions of the galaxy), our artificial nova simulations
may underestimate the overall completeness of our survey by
placing a greater fraction of the artificial novae in the
bright central regions of the galaxy where they are
generally more difficult to detect. This bias,
although we consider it to be small,
would have the effect of slightly overestimating
the nova rate determined earlier in our Monte Carlo simulations.

On the other hand, it is worth noting that 
M83 is an active
star-forming galaxy \citep[e.g., see][]{1999AJ....118..797C}, which
likely interacted $\sim$1-2~GYr ago with the nearby
metal-poor dwarf galaxy, NGC 5253, triggering
starbursts in both galaxies \citep{1980PASP...92..122V,1999AJ....118..797C}.
The starburst activity in M83 is concentrated in the spiral arms and
near the nucleus of the
galaxy, which is heavily shrouded in dust \citep{1991A&A...243..309G,
2008ASPC..390..144L}.
These are the same regions of the galaxy where the high surface brightness
and complex structure render the detection of novae difficult.
Thus, despite our careful nova searches,
the possibility that we may be missing some novae in these regions
of the galaxy cannot be definitively ruled out.
If such a putative population
of heavily-extincted novae exists, it would not be properly accounted for by
our artificial nova tests, the completeness would be overestimated, and
our simulations would then underestimate the true nova rate.

\subsection{The Observed Nova Spatial Distribution}

As discussed earlier, it was surprising to find
that the observed spatial distribution of
novae in M83 did not follow the background light of the galaxy.
To put this result into context,
it is instructive to compare the observed spatial distribution
of novae in M83 with similar galaxies for which nova populations
have been studied. Such galaxies include
the massive, late-type, and nearly face-on spirals
M51 and M101 (morphological types SA(s)bc and SAB(rs)cd, respectively),
the relatively low-mass, late-type systems
NGC~2403 and M33 (SAB(s)cd and SA(s)cd, respectively),
and M81, a relatively early-type
SA(s)ab spiral with a prominent bulge component.
Unfortunately, as we explore below, a review of these studies
does not suggest a simple association between
the morphological type of the galaxy and the
degree to which the nova distribution follows
the light of the host galaxy.

In the case of the face-on, grand-design spiral
M101 \citep{2000ApJ...530..193S,2008ApJ...686.1261C},
the nova density appears to track the
background light remarkably well (K-S, p=0.94), while for
M33 \citep{2004ApJ...612..867W, 2012ApJ...752..156S},
the cumulative distributions of the novae and the background light
were only marginally consistent (K-S, p=0.4).
Similar to our result for M83, the spatial distributions of novae in
both M51 \citep{2000ApJ...530..193S} and
NGC~2403 \citep{2012ApJ...760...13F}, were also found
to be more spatially extended than the background light
(with 97\% and 75\% confidence, respectively).
In theses cases, however, observational
incompleteness in the central region of M51 and the spiral arms of NGC~2403
could possibly explain the discrepancy.

The situation with regard to M81 is more complex. The spatial distribution
of novae in this galaxy has been studied extensively by a number of groups,
all of whom have come to somewhat different conclusions about how
the nova distribution compares with the galaxy light.
Based on a total of 15 and 12 detected novae, respectively,
\citet{1993AAS...182.8505M} and \citet{2004AJ....127..816N}
argued that the M81 novae most
closely follow the galaxy's bulge light, as was found in several
studies of the nova population in M31
\citep{1987ApJ...318..520C,2000ApJ...530..193S,2006MNRAS.369..257D}.
More recently, however, \citet{2008A&A...492..301H} analyzed
a much larger sample of novae (49) than either of the previous studies,
finding that the nova density distribution
best matches the overall (bulge + disk) background galaxy light. Given that
M81 has a prominent bulge component that dominates the overall light
in regions where most novae are found, the apparent discrepancy between these
studies may be
a distinction without much of a difference. In reviewing
the earlier studies in more detail, it appears that neither the
Moses \& Shafter
nor the Neill \& Shara data are inconsistent with the hypothesis that
the novae follow the overall light.

The only M81 nova study that reached
a very different conclusion is that of \citet{1999PASP..111.1367S},
who analyzed a sample of 23 novae discovered nearly a half century earlier
on Palomar photographic plates taken between 1950 and 1955. They found a
significant (disk) population of novae in the outer regions of the galaxy that
resulted in a poor fit of the full nova distribution to the overall
galaxy light
(and clearly a worse fit to the bulge light). The poor fit to the overall
light persisted even after the authors attempted to correct their data
for missing novae in the inner bulge region of the galaxy.
It is not obvious how to reconcile these early photographic
data with subsequent CCD
studies other than to posit that perhaps an even larger number
of bulge novae than expected
were missed due to the difficulty in detecting them on photographic
plates when projected against the bright background of the
galaxy's bulge.

Taken together, the best evidence currently available
suggests that the nova distribution in M81 likely follows the overall
background light of the galaxy, with novae belonging to both the bulge
and disk populations. In the case of most other galaxies, the situation
remains less clear. A combination of small-number statistics, spatial
variations in extinction (especially in late-type spiral galaxies),
and limited spatial coverage (e.g., in M31) have all conspired to
make it difficult to differentiate a real spatial (stellar
population) variation from biases caused by observational incompleteness.
As is often the case, more data will be required
before we have a full and complete understanding of how the nova properties,
including their specific rates, vary with the underlying stellar population.

\section{Summary} \label{sec:sum}

The principal results of our M83 nova survey are as follows:

\textbf{(1)} We have conducted an imaging survey of the
SAB(s)c starburst galaxy M83 and discovered a total of 19
novae over the course of our seven-year survey.

\textbf{(2)} After correcting for the survey's limiting magnitude
and spatial and temporal coverage, we find an overall
nova rate of $R_{M83} = 19^{+5}_{-3}$~yr$^{-1}$ for the galaxy.

\textbf{(3)} Adopting an integrated $K$-band magnitude
of $K_{M83} = 4.62\pm0.03$ from the 2MASS survey, and a distance
modulus for M83 of $(m-M)_0 = 28.34\pm0.14$, we find
the absolute magnitude of M83 in the $K$-band is $M_K=-23.72$, which
corresponds to a luminosity of $6.32\times10^{10}$~L$_{\odot,K}$.
The luminosity-specific nova rate of M83 is then found to be
$\nu_K = 3.0^{+0.9}_{-0.6}\times10^{-10}$~yr$^{-1}~L_{\odot,K}^{-1}$.

\textbf{(4)} The value of $\nu_K$ for M83 is typical of those found for other
galaxies with measured nova rates, and, in agreement with our earlier studies,
no compelling evidence is found for a variation of $\nu_K$ with galaxy
color, or Hubble type.

\textbf{(5)} Our survey has enabled us to measure light curves
for a total of 18 of the 19 novae discovered in M83. Of these,
the peak brightnesses and fade rates
($t_2$ times) could be measured for 17 of the novae. These data
show that the most luminous novae we observed
in M83 generally faded the fastest from maximum light in accordance
with the canonical MMRD relation.

\textbf{(6)}
We have found the spatial distribution of novae in M83 to be more
extended than the background galaxy light suggesting that they are
predominately associated with the disk population of the galaxy. In addition,
the M83 novae appear, on average, to reach higher luminosities and to
evolve more quickly compared with novae in M31, which are predominately
associated with that galaxy's bulge.
These findings are consistent with the claim made by
\citet{1990LNP...369...34D},
\citet{1998ApJ...506..818D}, and others that there are two
populations of novae, a ``disk" population characterized
by generally brighter and faster novae, and a ``bulge" population,
characterized by novae that are typically slower and less luminous.

\begin{acknowledgments}
We thank the anonymous
referee for their valuable comments and suggestions that have
helped us to improve the focus and presentation of this study.
This work is based on data collected with the Danish 1.54-m telescope
at the ESO La Silla Observatory.
We thank J. Vra\v{s}til, E. Kortusov\'a, L. Pilar\v{c}\'ik, and
V. Votruba for acquiring some M83 images,
J. L. Prieto for providing the M83
image taken with MEGACAM on the 6.5-m Magellan II telescope, and
W. Burris for assisting with the artificial nova tests
used to assess the limiting magnitude of our survey.
The work of K.H., H.K., L.K., M.S., P.K., and P.\v{S}.
was supported by the project RVO:67985815.
The research of M.W. and P.Z. was supported by the
project {\sc Progres Q47 Physics} of the Charles University in Prague.
M.S. acknowledges the support by Inter-transfer grant no LTT-20015.
\end{acknowledgments}

%\vspace{5mm}
\facilities{1.54-m Danish Telescope, La Silla}

\software{APHOT \citep{ 1994ExA.....5..375P}, IRAF \citep{1986SPIE..627..733T}, ISIS \citep{1998ApJ...503..325A}  
          }

\bibliography{m83novae}{}

\begin{thebibliography}{}
\expandafter\ifx\csname natexlab\endcsname\relax\def\natexlab#1{#1}\fi
\providecommand{\url}[1]{\href{#1}{#1}}
\providecommand{\dodoi}[1]{doi:~\href{http://doi.org/#1}{\nolinkurl{#1}}}
\providecommand{\doeprint}[1]{\href{http://ascl.net/#1}{\nolinkurl{http://ascl.net/#1}}}
\providecommand{\doarXiv}[1]{\href{https://arxiv.org/abs/#1}{\nolinkurl{https://arxiv.org/abs/#1}}}

\bibitem[{{Alard} \& {Lupton}(1998)}]{1998ApJ...503..325A}
{Alard}, C., \& {Lupton}, R.~H. 1998, \apj, 503, 325, \dodoi{10.1086/305984}

\bibitem[{{Bhardwaj} {et~al.}(2019){Bhardwaj}, {Kanbur}, {He}, {Rejkuba},
  {Matsunaga}, {de Grijs}, {Sharma}, {Singh}, {Baug}, {Ngeow}, \&
  {Ou}}]{2019ApJ...884...20B}
{Bhardwaj}, A., {Kanbur}, S., {He}, S., {et~al.} 2019, \apj, 884, 20,
  \dodoi{10.3847/1538-4357/ab38c2}

\bibitem[{{Calzetti} {et~al.}(1999){Calzetti}, {Conselice}, {Gallagher}, \&
  {Kinney}}]{1999AJ....118..797C}
{Calzetti}, D., {Conselice}, C.~J., {Gallagher}, John~S., I., \& {Kinney},
  A.~L. 1999, \aj, 118, 797, \dodoi{10.1086/300972}

\bibitem[{{Capaccioli} {et~al.}(1989){Capaccioli}, {Della Valle}, {D'Onofrio},
  \& {Rosino}}]{1989AJ.....97.1622C}
{Capaccioli}, M., {Della Valle}, M., {D'Onofrio}, M., \& {Rosino}, L. 1989,
  \aj, 97, 1622, \dodoi{10.1086/115104}

\bibitem[{{Ciardullo} {et~al.}(1987){Ciardullo}, {Ford}, {Neill}, {Jacoby}, \&
  {Shafter}}]{1987ApJ...318..520C}
{Ciardullo}, R., {Ford}, H.~C., {Neill}, J.~D., {Jacoby}, G.~H., \& {Shafter},
  A.~W. 1987, \apj, 318, 520, \dodoi{10.1086/165388}

\bibitem[{{Ciardullo} {et~al.}(1990){Ciardullo}, {Ford}, {Williams}, {Tamblyn},
  \& {Jacoby}}]{1990AJ.....99.1079C}
{Ciardullo}, R., {Ford}, H.~C., {Williams}, R.~E., {Tamblyn}, P., \& {Jacoby},
  G.~H. 1990, \aj, 99, 1079, \dodoi{10.1086/115397}

\bibitem[{{Coelho} {et~al.}(2008){Coelho}, {Shafter}, \&
  {Misselt}}]{2008ApJ...686.1261C}
{Coelho}, E.~A., {Shafter}, A.~W., \& {Misselt}, K.~A. 2008, \apj, 686, 1261,
  \dodoi{10.1086/591517}

\bibitem[{{Cohen}(1985)}]{1985ApJ...292...90C}
{Cohen}, J.~G. 1985, \apj, 292, 90, \dodoi{10.1086/163135}

\bibitem[{{Darnley} {et~al.}(2014){Darnley}, {Williams}, {Bode}, {Henze},
  {Ness}, {Shafter}, {Hornoch}, \& {Votruba}}]{2014A&A...563L...9D}
{Darnley}, M.~J., {Williams}, S.~C., {Bode}, M.~F., {et~al.} 2014, \aap, 563,
  L9, \dodoi{10.1051/0004-6361/201423411}

\bibitem[{{Darnley} {et~al.}(2006){Darnley}, {Bode}, {Kerins}, {Newsam}, {An},
  {Baillon}, {Belokurov}, {Calchi Novati}, {Carr}, {Cr{\'e}z{\'e}}, {Evans},
  {Giraud-H{\'e}raud}, {Gould}, {Hewett}, {Jetzer}, {Kaplan},
  {Paulin-Henriksson}, {Smartt}, {Tsapras}, \& {Weston}}]{2006MNRAS.369..257D}
{Darnley}, M.~J., {Bode}, M.~F., {Kerins}, E., {et~al.} 2006, \mnras, 369, 257,
  \dodoi{10.1111/j.1365-2966.2006.10297.x}

\bibitem[{{Darnley} {et~al.}(2017){Darnley}, {Hounsell}, {Godon}, {Perley},
  {Henze}, {Kuin}, {Williams}, {Williams}, {Bode}, {Harman}, {Hornoch}, {Link},
  {Ness}, {Ribeiro}, {Sion}, {Shafter}, \& {Shara}}]{2017ApJ...849...96D}
{Darnley}, M.~J., {Hounsell}, R., {Godon}, P., {et~al.} 2017, \apj, 849, 96,
  \dodoi{10.3847/1538-4357/aa9062}

\bibitem[{{de Vaucouleurs} {et~al.}(1991){de Vaucouleurs}, {de Vaucouleurs},
  {Corwin}, {Buta}, {Paturel}, \& {Fouque}}]{1991rc3..book.....D}
{de Vaucouleurs}, G., {de Vaucouleurs}, A., {Corwin}, Herold~G., J., {et~al.}
  1991, {Third Reference Catalogue of Bright Galaxies}

\bibitem[{{Della Valle} \& {Izzo}(2020)}]{2020A&ARv..28....3D}
{Della Valle}, M., \& {Izzo}, L. 2020, \aapr, 28, 3,
  \dodoi{10.1007/s00159-020-0124-6}

\bibitem[{{della Valle} \& {Livio}(1995)}]{1995ApJ...452..704D}
{della Valle}, M., \& {Livio}, M. 1995, \apj, 452, 704, \dodoi{10.1086/176342}

\bibitem[{{Della Valle} \& {Livio}(1998)}]{1998ApJ...506..818D}
{Della Valle}, M., \& {Livio}, M. 1998, \apj, 506, 818, \dodoi{10.1086/306275}

\bibitem[{{della Valle} {et~al.}(1994){della Valle}, {Rosino}, {Bianchini}, \&
  {Livio}}]{1994A&A...287..403D}
{della Valle}, M., {Rosino}, L., {Bianchini}, A., \& {Livio}, M. 1994, \aap,
  287, 403

\bibitem[{{Doggett} \& {Branch}(1985)}]{1985AJ.....90.2303D}
{Doggett}, J.~B., \& {Branch}, D. 1985, \aj, 90, 2303, \dodoi{10.1086/113934}

\bibitem[{{Downes} \& {Duerbeck}(2000)}]{2000AJ....120.2007D}
{Downes}, R.~A., \& {Duerbeck}, H.~W. 2000, \aj, 120, 2007,
  \dodoi{10.1086/301551}

\bibitem[{{Duerbeck}(1987)}]{1987SSRv...45....1D}
{Duerbeck}, H.~W. 1987, \ssr, 45, 1, \dodoi{10.1007/BF00187826}

\bibitem[{{Duerbeck}(1990)}]{1990LNP...369...34D}
---. 1990, {Galactic Distribution and Outburst Frequency of Classical Novae},
  ed. A.~{Cassatella} \& R.~{Viotti}, Vol. 369, 34,
  \dodoi{10.1007/3-540-53500-4\_90}

\bibitem[{{Franck} {et~al.}(2012){Franck}, {Shafter}, {Hornoch}, \&
  {Misselt}}]{2012ApJ...760...13F}
{Franck}, J.~R., {Shafter}, A.~W., {Hornoch}, K., \& {Misselt}, K.~A. 2012,
  \apj, 760, 13, \dodoi{10.1088/0004-637X/760/1/13}

\bibitem[{{Gallais} {et~al.}(1991){Gallais}, {Rouan}, {Lacombe}, {Tiphene}, \&
  {Vauglin}}]{1991A&A...243..309G}
{Gallais}, P., {Rouan}, D., {Lacombe}, F., {Tiphene}, D., \& {Vauglin}, I.
  1991, \aap, 243, 309

\bibitem[{{G{\"u}th} {et~al.}(2010){G{\"u}th}, {Shafter}, \&
  {Misselt}}]{2010ApJ...720.1155G}
{G{\"u}th}, T., {Shafter}, A.~W., \& {Misselt}, K.~A. 2010, \apj, 720, 1155,
  \dodoi{10.1088/0004-637X/720/2/1155}

\bibitem[{{Hornoch}(2013{\natexlab{a}})}]{2013ATel.4723....1H}
{Hornoch}, K. 2013{\natexlab{a}}, The Astronomer's Telegram, 4723, 1

\bibitem[{{Hornoch}(2013{\natexlab{b}})}]{2013ATel.4732....1H}
---. 2013{\natexlab{b}}, The Astronomer's Telegram, 4732, 1

\bibitem[{{Hornoch} \& {Kucakova}(2018)}]{2018ATel11443....1H}
{Hornoch}, K., \& {Kucakova}, H. 2018, The Astronomer's Telegram, 11443, 1

\bibitem[{{Hornoch} \& {Kucakova}(2019)}]{2019ATel12539....1H}
---. 2019, The Astronomer's Telegram, 12539, 1

\bibitem[{{Hornoch} {et~al.}(2019){Hornoch}, {Kucakova}, \&
  {Kurfurst}}]{2019ATel12564....1H}
{Hornoch}, K., {Kucakova}, H., \& {Kurfurst}, P. 2019, The Astronomer's
  Telegram, 12564, 1

\bibitem[{{Hornoch} \& {Paunzen}(2018)}]{2018ATel11240....1H}
{Hornoch}, K., \& {Paunzen}, E. 2018, The Astronomer's Telegram, 11240, 1

\bibitem[{{Hornoch} {et~al.}(2008){Hornoch}, {Scheirich}, {Garnavich},
  {Hameed}, \& {Thilker}}]{2008A&A...492..301H}
{Hornoch}, K., {Scheirich}, P., {Garnavich}, P.~M., {Hameed}, S., \& {Thilker},
  D.~A. 2008, \aap, 492, 301, \dodoi{10.1051/0004-6361:200809592}

\bibitem[{{Hornoch} {et~al.}(2013){Hornoch}, {Zasche}, \&
  {Wolf}}]{2013ATel.4747....1H}
{Hornoch}, K., {Zasche}, P., \& {Wolf}, M. 2013, The Astronomer's Telegram,
  4747, 1

\bibitem[{{Hubble}(1929)}]{1929ApJ....69..103H}
{Hubble}, E.~P. 1929, \apj, 69, 103, \dodoi{10.1086/143167}

\bibitem[{{Ivezi{\'c}} {et~al.}(2019){Ivezi{\'c}}, {Kahn}, {Tyson}, {Abel},
  {Acosta}, {Allsman}, {Alonso}, {AlSayyad}, {Anderson}, {Andrew}, {Angel},
  {Angeli}, {Ansari}, {Antilogus}, {Araujo}, {Armstrong}, {Arndt}, {Astier},
  {Aubourg}, {Auza}, {Axelrod}, {Bard}, {Barr}, {Barrau}, {Bartlett}, {Bauer},
  {Bauman}, {Baumont}, {Bechtol}, {Bechtol}, {Becker}, {Becla}, {Beldica},
  {Bellavia}, {Bianco}, {Biswas}, {Blanc}, {Blazek}, {Blandford}, {Bloom},
  {Bogart}, {Bond}, {Booth}, {Borgland}, {Borne}, {Bosch}, {Boutigny},
  {Brackett}, {Bradshaw}, {Brandt}, {Brown}, {Bullock}, {Burchat}, {Burke},
  {Cagnoli}, {Calabrese}, {Callahan}, {Callen}, {Carlin}, {Carlson},
  {Chandrasekharan}, {Charles-Emerson}, {Chesley}, {Cheu}, {Chiang}, {Chiang},
  {Chirino}, {Chow}, {Ciardi}, {Claver}, {Cohen-Tanugi}, {Cockrum}, {Coles},
  {Connolly}, {Cook}, {Cooray}, {Covey}, {Cribbs}, {Cui}, {Cutri}, {Daly},
  {Daniel}, {Daruich}, {Daubard}, {Daues}, {Dawson}, {Delgado}, {Dellapenna},
  {de Peyster}, {de Val-Borro}, {Digel}, {Doherty}, {Dubois},
  {Dubois-Felsmann}, {Durech}, {Economou}, {Eifler}, {Eracleous}, {Emmons},
  {Fausti Neto}, {Ferguson}, {Figueroa}, {Fisher-Levine}, {Focke}, {Foss},
  {Frank}, {Freemon}, {Gangler}, {Gawiser}, {Geary}, {Gee}, {Geha}, {Gessner},
  {Gibson}, {Gilmore}, {Glanzman}, {Glick}, {Goldina}, {Goldstein}, {Goodenow},
  {Graham}, {Gressler}, {Gris}, {Guy}, {Guyonnet}, {Haller}, {Harris},
  {Hascall}, {Haupt}, {Hernandez}, {Herrmann}, {Hileman}, {Hoblitt}, {Hodgson},
  {Hogan}, {Howard}, {Huang}, {Huffer}, {Ingraham}, {Innes}, {Jacoby}, {Jain},
  {Jammes}, {Jee}, {Jenness}, {Jernigan}, {Jevremovi{\'c}}, {Johns}, {Johnson},
  {Johnson}, {Jones}, {Juramy-Gilles}, {Juri{\'c}}, {Kalirai}, {Kallivayalil},
  {Kalmbach}, {Kantor}, {Karst}, {Kasliwal}, {Kelly}, {Kessler}, {Kinnison},
  {Kirkby}, {Knox}, {Kotov}, {Krabbendam}, {Krughoff}, {Kub{\'a}nek},
  {Kuczewski}, {Kulkarni}, {Ku}, {Kurita}, {Lage}, {Lambert}, {Lange},
  {Langton}, {Le Guillou}, {Levine}, {Liang}, {Lim}, {Lintott}, {Long},
  {Lopez}, {Lotz}, {Lupton}, {Lust}, {MacArthur}, {Mahabal}, {Mandelbaum},
  {Markiewicz}, {Marsh}, {Marshall}, {Marshall}, {May}, {McKercher}, {McQueen},
  {Meyers}, {Migliore}, {Miller}, {Mills}, {Miraval}, {Moeyens}, {Moolekamp},
  {Monet}, {Moniez}, {Monkewitz}, {Montgomery}, {Morrison}, {Mueller},
  {Muller}, {Mu{\~n}oz Arancibia}, {Neill}, {Newbry}, {Nief}, {Nomerotski},
  {Nordby}, {O'Connor}, {Oliver}, {Olivier}, {Olsen}, {O'Mullane}, {Ortiz},
  {Osier}, {Owen}, {Pain}, {Palecek}, {Parejko}, {Parsons}, {Pease},
  {Peterson}, {Peterson}, {Petravick}, {Libby Petrick}, {Petry},
  {Pierfederici}, {Pietrowicz}, {Pike}, {Pinto}, {Plante}, {Plate}, {Plutchak},
  {Price}, {Prouza}, {Radeka}, {Rajagopal}, {Rasmussen}, {Regnault}, {Reil},
  {Reiss}, {Reuter}, {Ridgway}, {Riot}, {Ritz}, {Robinson}, {Roby}, {Roodman},
  {Rosing}, {Roucelle}, {Rumore}, {Russo}, {Saha}, {Sassolas}, {Schalk},
  {Schellart}, {Schindler}, {Schmidt}, {Schneider}, {Schneider}, {Schoening},
  {Schumacher}, {Schwamb}, {Sebag}, {Selvy}, {Sembroski}, {Seppala}, {Serio},
  {Serrano}, {Shaw}, {Shipsey}, {Sick}, {Silvestri}, {Slater}, {Smith},
  {Smith}, {Sobhani}, {Soldahl}, {Storrie-Lombardi}, {Stover}, {Strauss},
  {Street}, {Stubbs}, {Sullivan}, {Sweeney}, {Swinbank}, {Szalay}, {Takacs},
  {Tether}, {Thaler}, {Thayer}, {Thomas}, {Thornton}, {Thukral}, {Tice},
  {Trilling}, {Turri}, {Van Berg}, {Vanden Berk}, {Vetter}, {Virieux},
  {Vucina}, {Wahl}, {Walkowicz}, {Walsh}, {Walter}, {Wang}, {Wang}, {Warner},
  {Wiecha}, {Willman}, {Winters}, {Wittman}, {Wolff}, {Wood-Vasey}, {Wu},
  {Xin}, {Yoachim}, \& {Zhan}}]{2019ApJ...873..111I}
{Ivezi{\'c}}, {\v{Z}}., {Kahn}, S.~M., {Tyson}, J.~A., {et~al.} 2019, \apj,
  873, 111, \dodoi{10.3847/1538-4357/ab042c}

\bibitem[{{Kasliwal} {et~al.}(2011){Kasliwal}, {Cenko}, {Kulkarni}, {Ofek},
  {Quimby}, \& {Rau}}]{2011ApJ...735...94K}
{Kasliwal}, M.~M., {Cenko}, S.~B., {Kulkarni}, S.~R., {et~al.} 2011, \apj, 735,
  94, \dodoi{10.1088/0004-637X/735/2/94}

\bibitem[{{Kato} {et~al.}(2014){Kato}, {Saio}, {Hachisu}, \&
  {Nomoto}}]{2014ApJ...793..136K}
{Kato}, M., {Saio}, H., {Hachisu}, I., \& {Nomoto}, K. 2014, \apj, 793, 136,
  \dodoi{10.1088/0004-637X/793/2/136}

\bibitem[{{Kuchinski} {et~al.}(2000){Kuchinski}, {Freedman}, {Madore},
  {Trewhella}, {Bohlin}, {Cornett}, {Fanelli}, {Marcum}, {Neff}, {O'Connell},
  {Roberts}, {Smith}, {Stecher}, \& {Waller}}]{2000ApJS..131..441K}
{Kuchinski}, L.~E., {Freedman}, W.~L., {Madore}, B.~F., {et~al.} 2000, \apjs,
  131, 441, \dodoi{10.1086/317371}

\bibitem[{{Landolt}(1992)}]{1992AJ....104..340L}
{Landolt}, A.~U. 1992, \aj, 104, 340, \dodoi{10.1086/116242}

\bibitem[{{Lundgren} {et~al.}(2008){Lundgren}, {Olofsson}, {Wiklind}, \&
  {Beck}}]{2008ASPC..390..144L}
{Lundgren}, A.~A., {Olofsson}, H., {Wiklind}, T., \& {Beck}, R. 2008, in
  Astronomical Society of the Pacific Conference Series, Vol. 390, Pathways
  Through an Eclectic Universe, ed. J.~H. {Knapen}, T.~J. {Mahoney}, \&
  A.~{Vazdekis}, 144

\bibitem[{{Mclaughlin}(1945)}]{1945PASP...57...69M}
{Mclaughlin}, D.~B. 1945, \pasp, 57, 69, \dodoi{10.1086/125689}

\bibitem[{{Moses} \& {Shafter}(1993)}]{1993AAS...182.8505M}
{Moses}, R.~N., \& {Shafter}, A.~W. 1993, in American Astronomical Society
  Meeting Abstracts, Vol. 182, American Astronomical Society Meeting Abstracts
  \#182, 85.05

\bibitem[{{Mr{\'o}z} {et~al.}(2016){Mr{\'o}z}, {Udalski}, {Poleski},
  {Soszy{\'n}ski}, {Szyma{\'n}ski}, {Pietrzy{\'n}ski}, {Wyrzykowski},
  {Ulaczyk}, {Koz{\l}owski}, {Pietrukowicz}, \&
  {Skowron}}]{2016ApJS..222....9M}
{Mr{\'o}z}, P., {Udalski}, A., {Poleski}, R., {et~al.} 2016, \apjs, 222, 9,
  \dodoi{10.3847/0067-0049/222/1/9}

\bibitem[{{Neill} \& {Shara}(2004)}]{2004AJ....127..816N}
{Neill}, J.~D., \& {Shara}, M.~M. 2004, \aj, 127, 816, \dodoi{10.1086/381484}

\bibitem[{{Nomoto}(1982)}]{1982ApJ...253..798N}
{Nomoto}, K. 1982, \apj, 253, 798, \dodoi{10.1086/159682}

\bibitem[{{Pravec} {et~al.}(1994){Pravec}, {Hudec}, {Sold{\'a}n}, {Sommer}, \&
  {Schenkl}}]{1994ExA.....5..375P}
{Pravec}, P., {Hudec}, R., {Sold{\'a}n}, J., {Sommer}, M., \& {Schenkl}, K.~H.
  1994, Experimental Astronomy, 5, 375, \dodoi{10.1007/BF01583708}

\bibitem[{{Prieto} \& {Morrell}(2013)}]{2013ATel.4734....1P}
{Prieto}, J.~L., \& {Morrell}, N. 2013, The Astronomer's Telegram, 4734, 1

\bibitem[{{Ritchey}(1917)}]{1917PASP...29..210R}
{Ritchey}, G.~W. 1917, \pasp, 29, 210, \dodoi{10.1086/122638}

\bibitem[{{Schaefer}(2018)}]{2018MNRAS.481.3033S}
{Schaefer}, B.~E. 2018, \mnras, 481, 3033, \dodoi{10.1093/mnras/sty2388}

\bibitem[{{Schlafly} \& {Finkbeiner}(2011)}]{2011ApJ...737..103S}
{Schlafly}, E.~F., \& {Finkbeiner}, D.~P. 2011, \apj, 737, 103,
  \dodoi{10.1088/0004-637X/737/2/103}

\bibitem[{{Shafter}(2019)}]{2019enhp.book.....S}
{Shafter}, A.~W. 2019, {Extragalactic Novae; A historical perspective},
  \dodoi{10.1088/2514-3433/ab2c63}

\bibitem[{{Shafter} {et~al.}(2000){Shafter}, {Ciardullo}, \&
  {Pritchet}}]{2000ApJ...530..193S}
{Shafter}, A.~W., {Ciardullo}, R., \& {Pritchet}, C.~J. 2000, \apj, 530, 193,
  \dodoi{10.1086/308349}

\bibitem[{{Shafter} {et~al.}(2014){Shafter}, {Curtin}, {Pritchet}, {Bode}, \&
  {Darnley}}]{2014ASPC..490...77S}
{Shafter}, A.~W., {Curtin}, C., {Pritchet}, C.~J., {Bode}, M.~F., \& {Darnley},
  M.~J. 2014, in Astronomical Society of the Pacific Conference Series, Vol.
  490, Stellar Novae: Past and Future Decades, ed. P.~A. {Woudt} \& V.~A.~R.~M.
  {Ribeiro}, 77.
\newblock \doarXiv{1307.2296}

\bibitem[{{Shafter} {et~al.}(2012){Shafter}, {Darnley}, {Bode}, \&
  {Ciardullo}}]{2012ApJ...752..156S}
{Shafter}, A.~W., {Darnley}, M.~J., {Bode}, M.~F., \& {Ciardullo}, R. 2012,
  \apj, 752, 156, \dodoi{10.1088/0004-637X/752/2/156}

\bibitem[{{Shafter} {et~al.}(2017){Shafter}, {Kundu}, \&
  {Henze}}]{2017RNAAS...1...11S}
{Shafter}, A.~W., {Kundu}, A., \& {Henze}, M. 2017, Research Notes of the
  American Astronomical Society, 1, 11, \dodoi{10.3847/2515-5172/aa9847}

\bibitem[{{Shafter} {et~al.}(2011){Shafter}, {Darnley}, {Hornoch},
  {Filippenko}, {Bode}, {Ciardullo}, {Misselt}, {Hounsell}, {Chornock}, \&
  {Matheson}}]{2011ApJ...734...12S}
{Shafter}, A.~W., {Darnley}, M.~J., {Hornoch}, K., {et~al.} 2011, \apj, 734,
  12, \dodoi{10.1088/0004-637X/734/1/12}

\bibitem[{{Shara} {et~al.}(1999){Shara}, {Sandage}, \&
  {Zurek}}]{1999PASP..111.1367S}
{Shara}, M.~M., {Sandage}, A., \& {Zurek}, D.~R. 1999, \pasp, 111, 1367,
  \dodoi{10.1086/316449}

\bibitem[{{Shara} {et~al.}(2016){Shara}, {Doyle}, {Lauer}, {Zurek}, {Neill},
  {Madrid}, {Miko{\l}ajewska}, {Welch}, \& {Baltz}}]{2016ApJS..227....1S}
{Shara}, M.~M., {Doyle}, T.~F., {Lauer}, T.~R., {et~al.} 2016, \apjs, 227, 1,
  \dodoi{10.3847/0067-0049/227/1/1}

\bibitem[{{Starrfield} {et~al.}(2016){Starrfield}, {Iliadis}, \&
  {Hix}}]{2016PASP..128e1001S}
{Starrfield}, S., {Iliadis}, C., \& {Hix}, W.~R. 2016, \pasp, 128, 051001,
  \dodoi{10.1088/1538-3873/128/963/051001}

\bibitem[{{Tang} {et~al.}(2014){Tang}, {Bildsten}, {Wolf}, {Li}, {Kong}, {Cao},
  {Cenko}, {De Cia}, {Kasliwal}, {Kulkarni}, {Laher}, {Masci}, {Nugent},
  {Perley}, {Prince}, \& {Surace}}]{2014ApJ...786...61T}
{Tang}, S., {Bildsten}, L., {Wolf}, W.~M., {et~al.} 2014, \apj, 786, 61,
  \dodoi{10.1088/0004-637X/786/1/61}

\bibitem[{{Tody}(1986)}]{1986SPIE..627..733T}
{Tody}, D. 1986, in Society of Photo-Optical Instrumentation Engineers (SPIE)
  Conference Series, Vol. 627, Instrumentation in astronomy VI, ed. D.~L.
  {Crawford}, 733, \dodoi{10.1117/12.968154}

\bibitem[{{Townsley} \& {Bildsten}(2005)}]{2005ApJ...628..395T}
{Townsley}, D.~M., \& {Bildsten}, L. 2005, \apj, 628, 395,
  \dodoi{10.1086/430594}

\bibitem[{{Tully} {et~al.}(2016){Tully}, {Courtois}, \&
  {Sorce}}]{2016AJ....152...50T}
{Tully}, R.~B., {Courtois}, H.~M., \& {Sorce}, J.~G. 2016, \aj, 152, 50,
  \dodoi{10.3847/0004-6256/152/2/50}

\bibitem[{{van den Bergh}(1980)}]{1980PASP...92..122V}
{van den Bergh}, S. 1980, \pasp, 92, 122, \dodoi{10.1086/130631}

\bibitem[{{Williams} \& {Shafter}(2004)}]{2004ApJ...612..867W}
{Williams}, S.~J., \& {Shafter}, A.~W. 2004, \apj, 612, 867,
  \dodoi{10.1086/422833}

\bibitem[{{Willmer}(2018)}]{2018ApJS..236...47W}
{Willmer}, C. N.~A. 2018, \apjs, 236, 47, \dodoi{10.3847/1538-4365/aabfdf}

\bibitem[{{Yungelson} {et~al.}(1997){Yungelson}, {Livio}, \&
  {Tutukov}}]{1997ApJ...481..127Y}
{Yungelson}, L., {Livio}, M., \& {Tutukov}, A. 1997, \apj, 481, 127,
  \dodoi{10.1086/304020}

\end{thebibliography}
\bibliographystyle{aasjournal}

\clearpage

\appendix

%\section{Tables 1 and 3 (these are machine readable in the published journal article)}

\startlongtable
\begin{deluxetable}{lccc}
\tablenum{1}
\tablecolumns{4}
\tablecaption{Log of Observations\label{tab:log}}
\tablehead{\colhead{UT Date} & \colhead{Julian Date} & \colhead{Limiting mag} & \\
\colhead{yr~mon~day} & \colhead{(2,450,000+)} & \colhead{($R$)} & \colhead{Notes\tablenotemark{a}}
}
\startdata
2012 12 12.361 &  6273.861 &  22.3 &             1 \cr
2012 12 18.368 &  6279.868 &  22.7 &             1 \cr
2012 12 22.352 &  6283.852 &  22.5 &             1 \cr
2012 12 23.345 &  6284.845 &  22.9 &             1 \cr
2012 12 28.314 &  6289.814 &  21.8 &             1 \cr
2013 01 02.313 &  6294.813 &  21.9 &             1 \cr
2013 01 08.374 &  6300.874 &  22.0 &             1 \cr
2013 01 09.370 &  6301.870 &  21.7 &             1 \cr
2013 01 10.374 &  6302.874 &  22.2 &             1 \cr
2013 01 11.367 &  6303.867 &  21.4 &             1 \cr
2013 01 12.374 &  6304.874 &  22.4 &             1 \cr
2013 01 14.384 &  6306.884 &  21.6 &             8 \cr
2013 01 15.389 &  6307.889 &  21.2 &             6 \cr
2013 01 17.368 &  6309.868 &  22.9 &             5 \cr
2013 01 18.360 &  6310.860 &  23.4 &             4 \cr
2013 01 19.387 &  6311.887 &  21.2 &             3 \cr
2013 01 22.373 &  6314.873 &  22.6 &             1 \cr
2013 01 23.387 &  6315.887 &  22.1 &             1 \cr
2013 01 26.382 &  6318.882 &  22.4 &             2 \cr
2013 01 28.389 &  6320.889 &  21.9 &             7 \cr
2013 01 29.378 &  6321.878 &  23.0 &             9 \cr
2013 01 31.362 &  6323.862 &  22.7 &             10 \cr
2013 02 01.308 &  6324.808 &  22.6 &             11 \cr
2013 02 03.327 &  6326.827 &  22.9 &             12 \cr
2013 02 05.390 &  6328.890 &  21.9 &             1 \cr
2013 02 07.397 &  6330.897 &  22.2 &             8 \cr
2013 02 08.395 &  6331.895 &  22.0 &             1 \cr
2013 02 09.381 &  6332.881 &  22.2 &             10 \cr
2013 02 11.387 &  6334.887 &  23.0 &             13 \cr
2013 02 12.389 &  6335.889 &  23.2 &             1 \cr
2013 02 13.397 &  6336.897 &  23.0 &             1 \cr
2013 02 14.398 &  6337.898 &  22.6 &             1 \cr
2013 02 16.396 &  6339.896 &  22.4 &             6 \cr
2013 02 18.388 &  6341.888 &  23.4 &             1 \cr
2013 02 19.385 &  6342.885 &  23.3 &             14 \cr
2013 02 20.405 &  6343.905 &  22.0 &             6 \cr
2013 02 22.362 &  6345.862 &  23.3 &             11 \cr
2013 02 24.349 &  6347.849 &  23.0 &             10 \cr
2013 02 25.386 &  6348.886 &  23.4 &             15 \cr
2013 03 01.338 &  6352.838 &  23.5 &             11 \cr
2013 03 04.397 &  6355.897 &  22.6 &             1 \cr
2013 03 06.401 &  6357.901 &  23.2 &             1 \cr
2013 03 08.404 &  6359.904 &  22.4 &             1 \cr
2013 03 09.404 &  6360.904 &  23.5 &             1 \cr
2013 03 14.244 &  6365.744 &  23.6 &             1 \cr
2013 03 24.377 &  6375.877 &  23.3 &             1 \cr
2012 04 01.240 &  6383.740 &  23.6 &             10 \cr
2013 04 05.364 &  6387.864 &  23.2 &             16 \cr
2013 04 06.249 &  6388.749 &  23.0 &             15 \cr
2013 04 07.301 &  6389.801 &  23.2 &             16 \cr
2013 04 08.360 &  6390.860 &  22.7 &             6 \cr
2013 04 09.125 &  6391.625 &  23.0 &             1 \cr
2013 04 11.422 &  6393.922 &  21.4 &             1 \cr
2013 04 16.420 &  6398.920 &  22.6 &             17 \cr
2013 04 17.104 &  6399.604 &  23.1 &             1 \cr
& & &\cr
2013 12 14.349 &  6640.849 &  22.7 &             10 \cr
2013 12 15.317 &  6641.817 &  22.2 &             11 \cr
2013 12 21.343 &  6647.843 &  22.7 &             19 \cr
2013 12 30.367 &  6656.867 &  22.4 &             10 \cr
2013 12 31.368 &  6657.868 &  23.2 &             1 \cr
2014 01 03.369 &  6660.869 &  22.6 &             1 \cr
2014 01 04.369 &  6661.869 &  22.8 &             1 \cr
2014 01 08.372 &  6665.872 &  22.3 &             1 \cr
2014 01 20.356 &  6677.856 &  23.5 &             11 \cr
2014 01 28.379 &  6685.879 &  23.0 &             1 \cr
2014 01 29.373 &  6686.873 &  23.6 &             1 \cr
2014 01 30.373 &  6687.873 &  23.1 &             1 \cr
2014 02 01.368 &  6689.868 &  22.7 &             15 \cr
2014 02 02.298 &  6690.798 &  23.5 &             12 \cr
2014 02 03.344 &  6691.844 &  23.0 &             16 \cr
2014 02 04.391 &  6692.891 &  23.8 &             1 \cr
2014 02 05.371 &  6693.871 &  23.3 &             1 \cr
2014 02 06.394 &  6694.894 &  22.7 &             1 \cr
2014 02 07.399 &  6695.899 &  23.0 &             1 \cr
2014 02 08.395 &  6696.895 &  22.9 &             6 \cr
2014 02 09.400 &  6697.900 &  22.4 &             6 \cr
2014 02 11.346 &  6699.846 &  23.4 &             10 \cr
2014 02 12.352 &  6700.852 &  23.2 &             11 \cr
2014 02 13.232 &  6701.732 &  22.9 &             11 \cr
2014 02 14.311 &  6702.811 &  23.0 &             11 \cr
2014 02 16.302 &  6704.802 &  22.9 &             12 \cr
2014 02 17.338 &  6705.838 &  22.7 &             10 \cr
2014 02 18.322 &  6706.822 &  22.5 &             11 \cr
2014 02 20.377 &  6708.877 &  22.8 &             10 \cr
2014 02 22.328 &  6710.828 &  23.2 &             19 \cr
2014 02 27.261 &  6715.761 &  23.5 &             19 \cr
2014 02 28.407 &  6716.907 &  22.5 &             1 \cr
2014 03 01.408 &  6717.908 &  22.2 &             1 \cr
2014 03 02.407 &  6718.907 &  22.4 &             1 \cr
2014 03 03.407 &  6719.907 &  22.4 &             1 \cr
2014 03 04.404 &  6720.904 &  23.0 &             1 \cr
2014 03 05.409 &  6721.909 &  22.6 &             1 \cr
2014 03 07.413 &  6723.913 &  22.1 &             6 \cr
2014 03 08.411 &  6724.911 &  22.1 &             1 \cr
2014 03 10.319 &  6726.819 &  23.5 &             12 \cr
2014 03 11.221 &  6727.721 &  23.8 &             11 \cr
2014 03 16.407 &  6732.907 &  22.8 &             1 \cr
2014 03 21.057 &  6737.557 &  23.0 &             1 \cr
2014 03 22.242 &  6738.742 &  23.4 &             16 \cr
2014 03 24.208 &  6740.708 &  23.5 &             19 \cr
2014 03 26.331 &  6742.831 &  23.1 &             1 \cr
2014 03 29.420 &  6745.920 &  23.1 &             1 \cr
2014 04 06.122 &  6753.622 &  22.8 &             12 \cr
2014 04 07.161 &  6754.661 &  23.5 &             15 \cr
2014 04 10.183 &  6757.683 &  23.0 &             20 \cr
& & &\cr
2014 12 24.327 &  7015.827 &  22.9 &             19 \cr
2014 12 25.355 &  7016.855 &  22.8 &             12 \cr
2014 12 27.344 &  7018.844 &  22.8 &             21 \cr
2015 01 02.346 &  7024.846 &  22.9 &             21 \cr
2015 01 09.351 &  7031.851 &  22.7 &             22 \cr
2015 01 10.362 &  7032.862 &  22.9 &             16 \cr
2015 01 11.352 &  7033.852 &  22.2 &             10 \cr
2015 01 12.345 &  7034.845 &  22.8 &             18 \cr
2015 01 13.339 &  7035.839 &  23.2 &             11 \cr
2015 01 15.379 &  7037.879 &  22.7 &             1 \cr
2015 01 16.375 &  7038.875 &  22.7 &             1 \cr
2015 01 19.381 &  7041.881 &  22.5 &             1 \cr
2015 01 20.383 &  7042.883 &  22.4 &             1 \cr
2015 01 22.349 &  7044.849 &  23.2 &             21 \cr
2015 01 25.388 &  7047.888 &  22.3 &             1 \cr
2015 01 27.390 &  7049.890 &  22.5 &             1 \cr
2015 01 28.374 &  7050.874 &  24.6 &             22 \cr
2015 01 31.375 &  7053.875 &  23.1 &             22 \cr
2015 02 01.307 &  7054.807 &  22.9 &             18 \cr
2015 02 03.376 &  7056.876 &  23.1 &             11 \cr
2015 02 04.304 &  7057.804 &  23.1 &             11 \cr
2015 02 06.375 &  7059.875 &  22.6 &             18 \cr
2015 02 10.329 &  7063.829 &  22.4 &             21 \cr
2015 02 12.394 &  7065.894 &  22.0 &             1 \cr
2015 02 13.394 &  7066.894 &  22.6 &             1 \cr
2015 02 14.396 &  7067.896 &  22.7 &             6 \cr
2015 02 17.402 &  7070.902 &  22.7 &             1 \cr
2015 02 18.402 &  7071.902 &  22.5 &             1 \cr
2015 02 22.378 &  7075.878 &  22.9 &             15 \cr
2015 03 03.342 &  7084.842 &  23.3 &             7 \cr
2015 03 05.288 &  7086.788 &  23.1 &             23 \cr
2015 03 12.407 &  7093.907 &  22.5 &             1 \cr
2015 03 13.414 &  7094.914 &  21.9 &             1 \cr
2015 03 14.412 &  7095.912 &  22.2 &             1 \cr
2015 03 17.413 &  7098.913 &  22.1 &             1 \cr
2015 03 19.416 &  7100.916 &  21.9 &             1 \cr
2015 03 20.409 &  7101.909 &  20.7 &             6 \cr
2015 03 28.422 &  7109.922 &  21.5 &             1 \cr
2015 03 31.328 &  7112.828 &  22.9 &             23 \cr
2015 04 01.321 &  7113.821 &  22.8 &             23 \cr
2015 04 08.415 &  7120.915 &  22.4 &             24 \cr
& & &\cr
2015 12 15.366 &  7371.866 &  21.2 &             1 \cr
2015 12 26.319 &  7382.819 &  22.4 &             21 \cr
2016 01 02.322 &  7389.822 &  22.8 &             12 \cr
2016 01 11.374 &  7398.874 &  22.4 &             1 \cr
2016 01 12.371 &  7399.871 &  22.8 &             1 \cr
2016 01 18.365 &  7405.865 &  23.2 &             21 \cr
2016 02 01.391 &  7419.891 &  22.5 &             1 \cr
2016 02 16.381 &  7434.881 &  23.4 &             23 \cr
2016 02 18.391 &  7436.891 &  23.9 &             18 \cr
2016 02 19.398 &  7437.898 &  23.5 &             16 \cr
2016 02 22.387 &  7440.887 &  23.2 &             16 \cr
2016 02 24.406 &  7442.906 &  21.2 &             21 \cr
2016 02 25.361 &  7443.861 &  22.5 &             10 \cr
2016 02 28.401 &  7446.901 &  22.1 &             19 \cr
2016 03 02.216 &  7449.716 &  22.5 &             19 \cr
2016 03 03.384 &  7450.884 &  23.6 &             1 \cr
2016 03 04.409 &  7451.909 &  22.9 &             1 \cr
2016 03 06.395 &  7453.895 &  23.8 &             6 \cr
2016 03 09.410 &  7456.910 &  22.1 &             1 \cr
2016 03 11.401 &  7458.901 &  23.3 &             21 \cr
2016 03 15.415 &  7462.915 &  22.5 &             1 \cr
2016 03 17.164 &  7464.664 &  22.9 &             18 \cr
2016 03 18.408 &  7465.908 &  23.2 &             25 \cr
2016 03 26.120 &  7473.620 &  22.6 &             26 \cr
2016 03 31.408 &  7478.908 &  22.9 &             16 \cr
& & &\cr
2017 12 23.363 &  8110.863 &  22.9 &             1 \cr
2018 01 28.263 &  8146.763 &  23.3 &             21 \cr
2018 01 29.278 &  8147.778 &  22.9 &             21 \cr
2018 01 30.244 &  8148.744 &  22.2 &             12 \cr
2018 02 03.360 &  8152.860 &  22.8 &             27 \cr
2018 02 06.220 &  8155.720 &  22.0 &             10 \cr
2018 02 07.259 &  8156.759 &  22.6 &             21 \cr
2018 02 15.397 &  8164.897 &  23.0 &             12 \cr
2018 02 26.233 &  8175.733 &  23.1 &             27 \cr
2018 02 27.208 &  8176.708 &  23.7 &             27 \cr
2018 02 28.200 &  8177.700 &  22.3 &             27 \cr
2018 03 13.395 &  8190.895 &  23.7 &             1 \cr
2018 03 15.415 &  8192.515 &  23.3 &             1 \cr
2018 03 17.409 &  8194.909 &  22.8 &             21 \cr
2018 03 19.417 &  8196.917 &  21.9 &             1 \cr
2018 03 22.417 &  8199.917 &  21.2 &             1 \cr
2018 03 26.207 &  8203.707 &  23.1 &             10 \cr
2018 03 28.415 &  8205.915 &  23.1 &             21 \cr
& & &\cr
2019 01 12.372 &  8495.872 &  22.2 &             1 \cr
2019 01 30.388 &  8513.888 &  21.9 &             27 \cr
2019 02 04.394 &  8518.894 &  21.8 &             27 \cr
2019 02 06.394 &  8520.894 &  22.1 &             27 \cr
2019 02 18.399 &  8532.899 &  22.2 &             27 \cr
2019 02 19.401 &  8533.901 &  22.6 &             27 \cr
2019 03 01.406 &  8543.906 &  22.7 &             1 \cr
2019 03 02.282 &  8544.782 &  23.5 &             28 \cr
2019 03 04.404 &  8546.904 &  22.7 &             1 \cr
2019 03 05.405 &  8547.905 &  23.0 &             1 \cr
2019 03 06.405 &  8548.905 &  22.9 &             1 \cr
2019 03 10.386 &  8552.886 &  22.8 &             1 \cr
2019 03 12.387 &  8554.887 &  23.0 &             1 \cr
2019 03 13.406 &  8555.906 &  22.9 &             1 \cr
2019 03 14.363 &  8556.863 &  23.7 &             1 \cr
2019 03 21.245 &  8563.745 &  22.7 &             27 \cr
\enddata
\tablenotetext{a}{All observations were made with the 1.54-m Danish Telescope at La Silla. The observers were:
(1)           K. Hornoch,
(2)   K. Hornoch and V. Votruba,
(3)  K. Hornoch and P. Ku\v{s}nir\'ak,
(4)      K. Hornoch and M. Wolf,
(5)     K. Hornoch and P. Zasche,
(6)          A. Gal\'ad,
(7)         L. Kotkov\'a,
(8)       P. Ku\v{s}nir\'ak,
(9)    K. Hornoch and P. \v{S}koda,
(10)         P. Zasche,
(11)          M. Wolf,
(12)         M. Zejda,
(13)     K. Hornoch and M. Zejda,
(14)    K. Hornoch and A. Gal\'ad,
(15)         M. Skarka,
(16)        J. Li\v{s}ka,
(17)         V. Votruba,
(18)        J. Vra\v{s}til,
(19)         J. Jan\'ik,
(20)       J. Ben\'a\v{c}ek,
(21)         E. Paunzen,
(22)       L. Pilar\v{c}\'ik,
(23)         E. Kortusov\'a,
(24)     E. Paunzen and M. Zejda,
(25)  M. Wolf and L. Pilar\v{c}\'ik,
(26)       J. Jury\v{s}ek,
(27)       H. Ku\v{c}\'akov\'a,
(28)        P. Kurf\"{u}rst.}
\end{deluxetable}

\startlongtable
\begin{deluxetable}{lcccc}
\tablenum{3}
\tablecolumns{5}
\tablecaption{M83 Nova Photometry}
\tablehead{\colhead{(UT Date)} & \colhead{(JD - 2,450,000)} & \colhead{Mag} & \colhead{Unc.} & \colhead{Band}
}
\startdata
\cutinhead{2013-nova1 = 2013-01a}
2012 06 01.138 &  6079.638 & [22.6  &\dots&R   \cr
2012 12 12.361 &  6273.861 & [22.3  &\dots&R   \cr
2012 12 18.368 &  6279.868 & [22.7  &\dots&R   \cr
2012 12 23.345 &  6284.845 & [22.9  &\dots&R   \cr
2012 12 28.314 &  6289.814 &  21.0  &0.15 &R   \cr
2012 12 28.316 &  6289.816 &  21.1  &0.2  &I   \cr
2013 01 02.313 &  6294.813 &  21.6  &0.2  &R   \cr
2013 01 08.374 &  6300.874 &  19.92 &0.10 &R   \cr
2013 01 09.370 &  6301.870 &  19.90 &0.10 &R   \cr
2013 01 10.374 &  6302.874 &  19.46 &0.09 &R   \cr
2013 01 11.367 &  6303.867 &  19.46 &0.10 &R   \cr
2013 01 12.374 &  6304.874 &  19.67 &0.07 &R   \cr
2013 01 14.384 &  6306.884 &  20.4  &0.1  &R   \cr
2013 01 15.389 &  6307.889 &  20.8  &0.3  &R   \cr
2013 01 17.368 &  6309.868 &  21.1  &0.15 &R   \cr
2013 01 18.360 &  6310.860 &  21.0  &0.1  &R   \cr
2013 01 18.380 &  6310.880 &  20.9  &0.2  &I   \cr
2013 01 19.387 &  6311.887 &  20.8  &0.25 &R   \cr
2013 01 22.373 &  6314.873 &  21.3  &0.15 &R   \cr
2013 01 23.387 &  6315.887 &  21.3  &0.15 &R   \cr
2013 01 26.382 &  6318.882 &  21.7  &0.2  &R   \cr
2013 01 28.389 &  6320.889 &  21.4  &0.25 &R   \cr
2013 01 29.378 &  6321.878 &  21.8  &0.2  &R   \cr
2013 01 29.392 &  6321.892 &  21.5  &0.3  &I   \cr
2013 01 31.362 &  6323.862 &  22.2  &0.2  &R    \cr
2013 02 01.308 &  6324.808 &  22.1  &0.25 &R    \cr
2013 02 03.327 &  6326.827 &  22.3  &0.3  &R    \cr
2013 02 05.390 &  6328.890 &  21.9  &0.3  &R   \cr
2013 02 07.397 &  6330.897 &  22.2  &0.35 &R   \cr
2013 02 08.395 &  6331.895 &  21.9  &0.3  &R   \cr
2013 02 09.381 &  6332.881 &  22.2  &0.3  &R    \cr
2013 02 11.387 &  6334.887 &  22.6  &0.2  &R    \cr
2013 02 12.389 &  6335.889 &  22.7  &0.25 &R   \cr
2013 02 13.397 &  6336.897 &  22.5  &0.3  &R   \cr
2013 02 14.398 &  6337.898 &  22.6  &0.25 &R   \cr
2013 02 16.396 &  6339.896 &  22.2  &0.3  &R   \cr
2013 02 18.388 &  6341.888 &  22.9  &0.3  &R   \cr
2013 02 19.385 &  6342.885 &  22.8  &0.3  &R    \cr
2013 02 22.362 &  6345.862 &  22.9  &0.25 &R    \cr
2013 02 24.349 &  6347.849 &  23.5  &0.4  &R    \cr
2013 02 25.386 &  6348.886 & [23.4  &\dots&R    \cr
2013 03 01.338 &  6352.838 & [23.5  &\dots&R    \cr
\cutinhead{2013-nova2 = 2013-01b}
2012 06 01.138 &  6079.638 & [22.8  &\dots&R   \cr
2013 01 02.313 &  6294.813 & [22.7  &\dots&R   \cr
2013 01 08.374 &  6300.874 &  21.5  &0.2  &R   \cr
2013 01 09.370 &  6301.870 &  21.2  &0.2  &R   \cr
2013 01 10.374 &  6302.874 &  21.0  &0.15 &R   \cr
2013 01 11.367 &  6303.867 &  21.1  &0.3  &R   \cr
2013 01 12.374 &  6304.874 &  21.8  &0.25 &R   \cr
2013 01 14.384 &  6306.884 &  21.5  &0.3  &R   \cr
2013 01 18.360 &  6310.860 &  22.1  &0.35 &R   \cr
2013 01 22.373 &  6314.873 &  22.6  &0.35 &R   \cr
2013 01 26.382 &  6318.882 &  22.8  &0.3  &R   \cr
2013 01 29.378 &  6321.878 & [23.0  &\dots&R   \cr
\cutinhead{2013-nova3 = 2013-01c}
2012 06 01.138 &  6079.638 & [22.8  &\dots&R   \cr
2012 12 28.314 &  6289.814 & [22.6  &\dots&R   \cr
2012 12 28.316 &  6289.816 & [22.2  &\dots&I   \cr
2013 01 02.313 &  6294.813 &  19.85 &0.09 &R   \cr
2013 01 09.370 &  6301.870 &  21.0  &0.15 &R   \cr
2013 01 10.374 &  6302.874 &  20.7  &0.15 &R   \cr
2013 01 11.367 &  6303.867 &  20.7  &0.2  &R   \cr
2013 01 12.374 &  6304.874 &  21.0  &0.15 &R   \cr
2013 01 14.384 &  6306.884 &  21.3  &0.25 &R   \cr
2013 01 17.368 &  6309.868 &  21.0  &0.15 &R   \cr
2013 01 18.360 &  6310.860 &  21.2  &0.15 &R   \cr
2013 01 19.387 &  6311.887 &  21.1  &0.3  &R   \cr
2013 01 22.373 &  6314.873 &  21.4  &0.2  &R   \cr
2013 01 23.387 &  6315.887 &  21.4  &0.2  &R   \cr
2013 01 26.382 &  6318.882 &  21.6  &0.2  &R   \cr
2013 01 28.389 &  6320.889 &  21.7  &0.25 &R   \cr
2013 01 29.378 &  6321.878 &  21.7  &0.2  &R   \cr
2013 01 29.392 &  6321.892 &  21.6  &0.35 &I   \cr
2013 01 31.362 &  6323.862 &  22.1  &0.4  &R    \cr
2013 02 01.308 &  6324.808 &  22.0  &0.35 &R    \cr
2013 02 03.327 &  6326.827 &  22.0  &0.35 &R    \cr
2013 02 05.390 &  6328.890 &  22.0  &0.35 &R   \cr
2013 02 07.397 &  6330.897 &  21.8  &0.3  &R   \cr
2013 02 08.395 &  6331.895 &  22.0  &0.3  &R   \cr
2013 02 09.381 &  6332.881 &  21.7  &0.3  &R    \cr
2013 02 11.387 &  6334.887 &  21.9  &0.2  &R    \cr
2013 02 12.389 &  6335.889 &  21.9  &0.2  &R   \cr
2013 02 13.397 &  6336.897 &  22.2  &0.3  &R   \cr
2013 02 14.398 &  6337.898 &  22.3  &0.3  &R   \cr
2013 02 16.396 &  6339.896 &  22.3  &0.3  &R   \cr
2013 02 18.388 &  6341.888 &  22.2  &0.3  &R   \cr
2013 02 19.385 &  6342.885 &  22.4  &0.3  &R    \cr
2013 02 22.362 &  6345.862 &  22.5  &0.25 &R    \cr
2013 02 24.349 &  6347.849 &  22.8  &0.3  &R    \cr
2013 02 25.386 &  6348.886 &  22.4  &0.25 &R    \cr
2013 03 01.338 &  6352.838 &  22.9  &0.3  &R    \cr
2013 03 04.397 &  6355.897 &  22.2  &0.3  &R   \cr
2013 03 06.401 &  6357.901 &  22.5  &0.25 &R   \cr
2013 03 09.404 &  6360.904 &  23.5  &0.4  &R   \cr
2013 03 14.244 &  6365.744 &  23.1  &0.35 &R   \cr
2013 03 24.377 &  6375.877 &  22.7  &0.3  &R   \cr
2013 04 06.249 &  6388.749 &  23.0  &0.4  &R    \cr
2013 04 17.104 &  6399.604 & [23.3  &\dots&R   \cr
\cutinhead{2013-nova4 = 2013-01d}
2013 01 14.384 &  6306.884 & [21.9  &\dots&R   \cr
2013 01 15.389 &  6307.889 & [21.5  &\dots&R   \cr
2013 01 18.360 &  6310.860 &  22.5  &0.25 &R   \cr
2013 01 18.380 &  6310.880 & [22.6  &\dots&I   \cr
2013 01 26.382 &  6318.882 &  21.2  &0.1  &R   \cr
2013 01 29.378 &  6321.878 &  21.06 &0.07 &R   \cr
2013 01 29.392 &  6321.892 &  20.8  &0.15 &I   \cr
2013 01 31.362 &  6323.862 &  21.05 &0.07 &R    \cr
2013 02 01.308 &  6324.808 &  20.91 &0.08 &R    \cr
2013 02 03.327 &  6326.827 &  20.89 &0.08 &R    \cr
2013 02 08.395 &  6331.895 &  20.92 &0.10 &R   \cr
2013 02 11.387 &  6334.887 &  20.96 &0.07 &R    \cr
2013 02 12.389 &  6335.889 &  21.15 &0.08 &R   \cr
2013 02 13.397 &  6336.897 &  21.43 &0.10 &R   \cr
2013 02 14.398 &  6337.898 &  21.2  &0.1  &R   \cr
2013 02 18.388 &  6341.888 &  21.5  &0.1  &R   \cr
2013 02 19.385 &  6342.885 &  21.7  &0.15 &R    \cr
2013 02 25.386 &  6348.886 &  21.8  &0.15 &R    \cr
2013 03 01.338 &  6352.838 &  22.0  &0.1  &R    \cr
2013 03 02.288 &  6353.788 &  22.0  &0.15 &I    \cr
2013 03 04.397 &  6355.897 &  21.8  &0.15 &R   \cr
2013 03 06.401 &  6357.901 &  21.9  &0.1  &R   \cr
2013 03 09.404 &  6360.904 &  22.3  &0.15 &R   \cr
2013 03 24.377 &  6375.877 &  23.0  &0.25 &R   \cr
2013 04 06.249 &  6388.749 &  22.8  &0.2  &R    \cr
2013 04 17.104 &  6399.604 &  23.5  &0.3  &R   \cr
2013 04 17.114 &  6399.614 &  22.7  &0.4  &I   \cr
2013 12 14.349 &  6640.849 & [23.2  &\dots&R    \cr
\cutinhead{2013-nova5 = 2013-04a}
2012 06 01.138 &  6079.638 & [23.0  &\dots&R   \cr
2012 04 01.240 &  6383.740 & [23.6  &\dots&R    \cr
2013 04 05.364 &  6387.864 &  20.9  &0.1  &R    \cr
2013 04 06.249 &  6388.749 &  20.20 &0.06 &R    \cr
2013 04 07.301 &  6389.801 &  20.33 &0.08 &R    \cr
2013 04 08.360 &  6390.860 &  20.25 &0.09 &R   \cr
2013 04 09.125 &  6391.625 &  20.48 &0.09 &R   \cr
2013 04 11.422 &  6393.922 &  21.1  &0.25 &R   \cr
2013 04 16.420 &  6398.920 &  21.6  &0.3  &R    \cr
2013 04 17.104 &  6399.604 &  21.5  &0.2  &R   \cr
2013 04 17.114 &  6399.614 &  20.8  &0.2  &I   \cr
2013 12 14.349 &  6640.849 & [22.7  &\dots&R    \cr
\cutinhead{2013-nova6 = 2013-03a}
2013 02 01.308 &  6324.808 & [23.4  &\dots&R    \cr
2013 02 03.327 &  6326.827 & [23.2  &\dots&R    \cr
2013 02 08.395 &  6331.895 &  22.2  &0.3  &R   \cr
2013 02 11.387 &  6334.887 &  21.9  &0.15 &R    \cr
2013 02 12.389 &  6335.889 &  22.1  &0.15 &R   \cr
2013 02 13.397 &  6336.897 &  21.9  &0.15 &R   \cr
2013 02 14.398 &  6337.898 &  21.8  &0.15 &R   \cr
2013 02 18.388 &  6341.888 &  21.74 &0.10 &R   \cr
2013 02 19.385 &  6342.885 &  21.67 &0.09 &R    \cr
2013 02 22.362 &  6345.862 &  21.54 &0.10 &R    \cr
2013 02 24.349 &  6347.849 &  21.7  &0.15 &R    \cr
2013 02 25.386 &  6348.886 &  21.8  &0.1  &R    \cr
2013 03 01.338 &  6352.838 &  21.72 &0.09 &R    \cr
2013 03 02.288 &  6353.788 &  21.0  &0.15 &I    \cr
2013 03 04.397 &  6355.897 &  21.7  &0.15 &R   \cr
2013 03 06.401 &  6357.901 &  21.9  &0.1  &R   \cr
2013 03 08.404 &  6359.904 &  21.9  &0.25 &R   \cr
2013 03 09.404 &  6360.904 &  22.1  &0.15 &R   \cr
2013 03 14.244 &  6365.744 &  21.9  &0.1  &R   \cr
2013 03 14.413 &  6365.913 &  21.0  &0.15 &I   \cr
2013 03 24.377 &  6375.877 &  21.9  &0.1  &R   \cr
2012 04 01.240 &  6383.740 &  22.1  &0.15 &R    \cr
2013 04 05.364 &  6387.864 &  22.4  &0.2  &R    \cr
2013 04 06.249 &  6388.749 &  22.0  &0.1  &R    \cr
2013 04 07.301 &  6389.801 &  22.6  &0.2  &R    \cr
2013 04 08.360 &  6390.860 &  22.4  &0.15 &R   \cr
2013 04 09.125 &  6391.625 &  22.3  &0.2  &R   \cr
2013 04 16.420 &  6398.920 &  22.3  &0.2  &R    \cr
2013 04 17.104 &  6399.604 &  22.4  &0.15 &R   \cr
2013 04 17.114 &  6399.614 &  21.3  &0.2  &I   \cr
2013 12 31.368 &  6657.868 & [23.2  &\dots&R   \cr
2014 03 11.221 &  6727.721 & [23.4  &\dots&R    \cr
2015 01 28.374 &  7050.874 & [23.8  &\dots&R    \cr
2016 02 18.391 &  7436.891 & [23.6  &\dots&R    \cr
2018 02 27.208 &  8176.708 & [23.7  &\dots&R    \cr
2019 03 14.363 &  8556.863 & [23.6  &\dots&R   \cr
\cutinhead{2014-nova1 = 2014-01a}
2014 01 09.371 &  6666.871 & [23.0  &\dots&R   \cr
2014 01 13.360 &  6670.860 &  19.60 &0.15 &R    \cr
2014 01 16.271 &  6673.771 &  19.42 &0.15 &R    \cr
2014 01 18.282 &  6675.782 &  20.0  &0.15 &R    \cr
2014 01 20.356 &  6677.856 &  20.30 &0.2  &R    \cr
2014 01 28.379 &  6685.879 &  22.1  &0.25 &R   \cr
2014 01 29.373 &  6686.873 &  23.3  &0.3  &R   \cr
2014 01 30.373 &  6687.873 &  22.8  &0.35 &R   \cr
2014 02 01.368 &  6689.868 &  23.5  &0.4  &R    \cr
2014 02 02.298 &  6690.798 &  24.3  &0.5  &R    \cr
2014 02 02.307 &  6690.807 & [22.7  &\dots&I    \cr
2014 02 03.344 &  6691.844 & [22.7  &\dots&R    \cr
2014 02 04.391 &  6692.891 & [23.8  &\dots&R   \cr
\cutinhead{2014-nova2 = 2014-01b}
2014 01 20.356 &  6677.856 & [23.5  &\dots&R    \cr
2014 01 28.379 &  6685.879 &  21.7  &0.15 &R   \cr
2014 01 29.373 &  6686.873 &  21.9  &0.15 &R   \cr
2014 01 30.373 &  6687.873 &  22.1  &0.15 &R   \cr
2014 02 01.368 &  6689.868 &  22.0  &0.15 &R    \cr
2014 02 02.298 &  6690.798 &  21.9  &0.15 &R    \cr
2014 02 02.307 &  6690.807 &  21.8  &0.2  &I    \cr
2014 02 03.344 &  6691.844 &  21.7  &0.2  &R    \cr
2014 02 04.391 &  6692.891 &  21.5  &0.15 &R   \cr
2014 02 05.371 &  6693.871 &  21.2  &0.15 &R   \cr
2014 02 06.394 &  6694.894 &  22.2  &0.2  &R   \cr
2014 02 07.399 &  6695.899 &  23.0  &0.4  &R   \cr
2014 02 08.395 &  6696.895 &  22.8  &0.4  &R   \cr
2014 02 09.400 &  6697.900 &  22.4  &0.5  &R   \cr
2014 02 11.346 &  6699.846 &  22.7  &0.25 &R    \cr
2014 02 12.352 &  6700.852 &  22.5  &0.25 &R    \cr
2014 02 13.232 &  6701.732 &  22.1  &0.25 &R    \cr
2014 02 14.311 &  6702.811 &  22.3  &0.3  &R    \cr
2014 02 16.302 &  6704.802 &  22.1  &0.3  &R    \cr
2014 02 17.338 &  6705.838 &  21.9  &0.35 &R    \cr
2014 02 18.322 &  6706.822 &  21.7  &0.35 &R    \cr
2014 02 20.377 &  6708.877 &  22.1  &0.25 &R    \cr
2014 02 22.328 &  6710.828 &  22.7  &0.3  &R    \cr
2014 02 27.261 &  6715.761 &  22.5  &0.3  &R    \cr
2014 02 28.407 &  6716.907 &  22.1  &0.5  &R   \cr
2014 03 01.408 &  6717.908 & [22.3  &\dots&R   \cr
2014 03 02.407 &  6718.907 &  22.7  &0.4  &R   \cr
2014 03 03.407 &  6719.907 &  22.6  &0.4  &R   \cr
2014 03 04.404 &  6720.904 &  22.9  &0.4  &R   \cr
2014 03 05.409 &  6721.909 &  22.9  &0.4  &R   \cr
2014 03 07.413 &  6723.913 & [22.1  &\dots&R   \cr
2014 03 08.411 &  6724.911 & [22.1  &\dots&R   \cr
2014 03 10.319 &  6726.819 &  24.1  &0.4  &R    \cr
2014 03 11.221 &  6727.721 &  23.3  &0.4  &R    \cr
2014 03 16.407 &  6732.907 & [22.8  &\dots&R   \cr
2014 03 21.057 &  6737.557 & [23.0  &\dots&R   \cr
2014 03 22.242 &  6738.742 & [23.8  &\dots&R    \cr
2014 03 24.208 &  6740.708 &  23.8  &0.4  &R    \cr
2014 03 26.331 &  6742.831 & [23.1  &\dots&R   \cr
2014 03 29.420 &  6745.920 & [23.1  &\dots&R   \cr
2014 04 06.122 &  6753.622 & [22.8  &\dots&R    \cr
2014 04 07.161 &  6754.661 & [23.5  &\dots&R    \cr
2014 04 10.183 &  6757.683 & [23.0  &\dots&R    \cr
\cutinhead{2014-nova3 = 2014-01c}
2014 01 20.356 &  6677.856 & [23.4  &\dots&R    \cr
2014 01 28.379 &  6685.879 &  22.7  &0.4  &R   \cr
2014 01 29.373 &  6686.873 &  22.0  &0.2  &R   \cr
2014 01 30.373 &  6687.873 &  20.9  &0.15 &R   \cr
2014 02 01.368 &  6689.868 &  19.4  &0.1  &R    \cr
2014 02 02.298 &  6690.798 &  18.94 &0.07 &R    \cr
2014 02 02.307 &  6690.807 &  18.7  &0.15 &I    \cr
2014 02 03.344 &  6691.844 &  18.89 &0.07 &R    \cr
2014 02 04.391 &  6692.891 &  19.1  &0.1  &R   \cr
2014 02 05.371 &  6693.871 &  19.3  &0.1  &R   \cr
2014 02 06.394 &  6694.894 &  20.15 &0.1  &R   \cr
2014 02 07.399 &  6695.899 &  20.6  &0.15 &R   \cr
2014 02 08.395 &  6696.895 &  21.1  &0.15 &R   \cr
2014 02 09.400 &  6697.900 &  21.4  &0.2  &R   \cr
2014 02 11.346 &  6699.846 &  21.6  &0.15 &R    \cr
2014 02 12.352 &  6700.852 &  21.7  &0.2  &R    \cr
2014 02 13.232 &  6701.732 &  21.5  &0.25 &R    \cr
2014 02 14.311 &  6702.811 &  21.5  &0.25 &R    \cr
2014 02 16.302 &  6704.802 &  21.6  &0.3  &R    \cr
2014 02 17.338 &  6705.838 &  22.0  &0.35 &R    \cr
2014 02 18.322 &  6706.822 &  21.8  &0.35 &R    \cr
2014 02 20.377 &  6708.877 &  21.9  &0.3  &R    \cr
2014 02 22.328 &  6710.828 &  22.1  &0.3  &R    \cr
2014 02 27.261 &  6715.761 &  22.4  &0.4  &R    \cr
2014 02 28.407 &  6716.907 &  22.1  &0.4  &R   \cr
2014 03 01.408 &  6717.908 & [22.2  &\dots&R   \cr
2014 03 02.407 &  6718.907 &  22.3  &0.4  &R   \cr
2014 03 03.407 &  6719.907 &  21.9  &0.4  &R   \cr
2014 03 04.404 &  6720.904 &  22.5  &0.4  &R   \cr
2014 03 05.409 &  6721.909 &  22.1  &0.4  &R   \cr
2014 03 07.413 &  6723.913 & [22.1  &\dots&R   \cr
2014 03 08.411 &  6724.911 & [22.1  &\dots&R   \cr
2014 03 10.319 &  6726.819 &  22.9  &0.35 &R    \cr
2014 03 11.221 &  6727.721 &  23.0  &0.4  &R    \cr
2014 03 16.407 &  6732.907 & [22.8  &\dots&R   \cr
2014 03 21.057 &  6737.557 & [23.0  &\dots&R   \cr
2014 03 22.242 &  6738.742 &  23.2  &0.4  &R    \cr
2014 03 24.208 &  6740.708 &  22.9  &0.4  &R    \cr
2014 03 26.331 &  6742.831 & [23.1  &\dots&R   \cr
2014 03 29.420 &  6745.920 & [22.6  &\dots&R   \cr
2014 04 06.122 &  6753.622 & [22.0  &\dots&R    \cr
2014 04 07.161 &  6754.661 & [23.0  &\dots&R    \cr
2014 04 10.183 &  6757.683 & [23.0  &\dots&R    \cr
\cutinhead{2014-nova4 = 2014-03a}
2014 03 03.407 &  6719.907 & [23.2  &\dots&R   \cr
2014 03 04.404 &  6720.904 &  21.2  &0.25 &R   \cr
2014 03 05.409 &  6721.909 &  19.75 &0.10 &R   \cr
2014 03 07.413 &  6723.913 &  19.8  &0.1  &R   \cr
2014 03 08.411 &  6724.911 &  19.95 &0.1  &R   \cr
2014 03 10.319 &  6726.819 &  20.06 &0.08 &R    \cr
2014 03 11.221 &  6727.721 &  20.5  &0.1  &R    \cr
2014 03 16.407 &  6732.907 &  22.7  &0.4  &R   \cr
2014 03 21.057 &  6737.557 &  23.0  &0.4  &R   \cr
2014 03 22.242 &  6738.742 &  22.9  &0.35 &R    \cr
2014 03 24.208 &  6740.708 &  22.7  &0.3  &R    \cr
2014 03 26.331 &  6742.831 & [23.1  &\dots&R   \cr
2014 03 29.420 &  6745.920 & [23.2  &\dots&R   \cr
2014 04 06.122 &  6753.622 & [23.1  &\dots&R    \cr
2014 04 07.161 &  6754.661 & [23.0  &\dots&R    \cr
2014 04 10.183 &  6757.683 & [22.4  &\dots&R    \cr
\cutinhead{2015-nova1 = 2015-01a}
2014 12 27.344 &  7018.844 & [23.3  &\dots&R    \cr
2015 01 02.346 &  7024.846 &  22.5  &0.1  &R    \cr
2015 01 09.351 &  7031.851 &  20.69 &0.08 &R    \cr
2015 01 10.362 &  7032.862 &  20.82 &0.09 &R    \cr
2015 01 11.352 &  7033.852 &  20.7  &0.1  &R    \cr
2015 01 11.354 &  7033.854 &  20.4  &0.15 &I    \cr
2015 01 12.345 &  7034.845 &  20.73 &0.05 &R    \cr
2015 01 13.339 &  7035.839 &  21.15 &0.06 &R    \cr
2015 01 15.379 &  7037.879 &  21.04 &0.08 &R   \cr
2015 01 16.375 &  7038.875 &  21.30 &0.09 &R   \cr
2015 01 19.381 &  7041.881 &  21.55 &0.12 &R   \cr
2015 01 20.383 &  7042.883 &  21.8  &0.15 &R   \cr
2015 01 22.349 &  7044.849 &  21.72 &0.08 &R    \cr
2015 01 25.388 &  7047.888 &  21.8  &0.15 &R   \cr
2015 01 27.390 &  7049.890 &  22.4  &0.25 &R   \cr
2015 01 28.374 &  7050.874 &  22.35 &0.09 &R    \cr
2015 01 31.375 &  7053.875 &  22.3  &0.15 &R    \cr
2015 02 01.307 &  7054.807 &  22.6  &0.2  &R    \cr
2015 02 03.376 &  7056.876 &  22.6  &0.2  &R    \cr
2015 02 04.304 &  7057.804 &  23.0  &0.35 &R    \cr
2015 02 06.375 &  7059.875 & [22.6  &\dots&R    \cr
\cutinhead{2015-nova2 = 2015-04a}
2015 04 01.321 &  7113.821 & [22.8  &\dots&R    \cr
2015 04 08.415 &  7120.915 &  21.4  &0.25 &R    \cr
\cutinhead{2016-nova1 = 2016-02a}
2016 02 18.391 &  7436.891 & [23.7  &\dots&R    \cr
2016 02 19.398 &  7437.898 &  22.9  &0.35 &R    \cr
2016 02 22.387 &  7440.887 &  21.14 &0.09 &R    \cr
2016 02 24.406 &  7442.906 &  21.0  &0.2  &R    \cr
2016 02 25.361 &  7443.861 &  20.84 &0.10 &R    \cr
2016 02 28.401 &  7446.901 &  20.8  &0.15 &R    \cr
2016 03 02.216 &  7449.716 &  21.4  &0.15 &R    \cr
2016 03 03.384 &  7450.884 &  21.44 &0.10 &R   \cr
2016 03 03.401 &  7450.901 &  21.5  &0.25 &I   \cr
2016 03 04.409 &  7451.909 &  21.4  &0.15 &R   \cr
2016 03 06.395 &  7453.895 &  21.53 &0.09 &R   \cr
2016 03 09.410 &  7456.910 &  21.6  &0.25 &R   \cr
2016 03 11.401 &  7458.901 &  22.1  &0.2  &R    \cr
2016 03 17.164 &  7464.664 &  22.2  &0.25 &R    \cr
2016 03 18.408 &  7465.908 &  22.4  &0.2  &R    \cr
2016 03 26.120 &  7473.620 & [22.6  &\dots&R    \cr
2016 03 31.408 &  7478.908 & [22.9  &\dots&R    \cr
\cutinhead{2016-nova2 = 2016-02b}
2016 02 18.391 &  7436.891 & [24.0  &\dots&R    \cr
2016 02 19.398 &  7437.898 & [23.5  &\dots&R    \cr
2016 02 22.387 &  7440.887 &  21.42 &0.09 &R    \cr
2016 02 24.406 &  7442.906 &  21.1  &0.2  &R    \cr
2016 02 25.361 &  7443.861 &  20.55 &0.10 &R    \cr
2016 02 28.401 &  7446.901 &  20.4  &0.15 &R    \cr
2016 03 02.216 &  7449.716 &  21.1  &0.15 &R    \cr
2016 03 03.384 &  7450.884 &  21.17 &0.09 &R   \cr
2016 03 03.401 &  7450.901 &  20.6  &0.1  &I   \cr
2016 03 04.409 &  7451.909 &  21.4  &0.15 &R   \cr
2016 03 06.395 &  7453.895 &  21.4  &0.1  &R   \cr
2016 03 09.410 &  7456.910 &  21.7  &0.25 &R   \cr
2016 03 11.401 &  7458.901 &  21.6  &0.15 &R    \cr
2016 03 17.164 &  7464.664 &  21.8  &0.15 &R    \cr
2016 03 18.408 &  7465.908 &  22.0  &0.15 &R    \cr
2016 03 26.120 &  7473.620 &  21.8  &0.3  &R    \cr
2016 03 31.408 &  7478.908 &  22.4  &0.3  &R    \cr
\cutinhead{2016-nova3 = 2016-03a}
2016 03 15.415 &  7462.915 & [22.5  &\dots&R   \cr
2016 03 17.164 &  7464.664 &  20.60 &0.09 &R    \cr
2016 03 18.408 &  7465.908 &  19.87 &0.08 &R    \cr
2016 03 26.120 &  7473.620 &  21.2  &0.1  &R    \cr
2016 03 31.408 &  7478.908 &  21.7  &0.2  &R    \cr
\cutinhead{2018-nova1 = 2018-01a}
2017 12 23.363 &  8110.863 & [22.9  &\dots&R   \cr
2018 01 28.263 &  8146.763 &  20.71 &0.08 &R    \cr
2018 01 29.278 &  8147.778 &  20.86 &0.09 &R    \cr
2018 01 29.290 &  8147.790 &  20.7  &0.1  &I    \cr
2018 01 30.244 &  8148.744 &  21.3  &0.1  &R    \cr
2018 02 03.360 &  8152.860 &  21.4  &0.1  &R    \cr
2018 02 06.220 &  8155.720 &  21.4  &0.3  &R    \cr
2018 02 07.259 &  8156.759 &  21.7  &0.15 &R    \cr
2018 02 15.397 &  8164.897 &  22.1  &0.25 &R    \cr
2018 02 26.233 &  8175.733 &  22.8  &0.3  &R    \cr
2018 02 27.208 &  8176.708 &  22.9  &0.3  &R    \cr
2018 02 28.200 &  8177.700 &  22.5  &0.3  &R    \cr
2018 03 13.395 &  8190.895 & [23.1  &\dots&R   \cr
2018 03 17.409 &  8194.909 & [23.1  &\dots&R    \cr
\cutinhead{2018-nova2 = 2018-02a}
2018 02 15.397 &  8164.897 & [23.2  &\dots&R    \cr
2018 02 26.233 &  8175.733 &  21.5  &0.15 &R    \cr
2018 02 27.208 &  8176.708 &  21.6  &0.15 &R    \cr
2018 02 28.200 &  8177.700 &  21.3  &0.25 &R    \cr
2018 03 13.395 &  8190.895 &  22.0  &0.2  &R   \cr
2018 03 15.415 &  8192.915 &  21.4  &0.3  &R   \cr
2018 03 17.409 &  8194.909 &  21.8  &0.25 &R    \cr
2018 03 19.417 &  8196.917 &  21.6  &0.25 &R   \cr
2018 03 21.417 &  8198.917 &  18.5  &0.2  &Ha  \cr
2018 03 26.207 &  8203.707 &  22.4  &0.2  &R    \cr
2018 03 28.415 &  8205.915 &  22.4  &0.5  &R    \cr
\cutinhead{2019-nova1 = 2019-02a}
2019 02 06.394 &  8520.894 & [22.1  &\dots&R    \cr
2019 02 18.399 &  8532.899 &  19.9  &0.1  &R    \cr
2019 02 19.401 &  8533.901 &  20.1  &0.15 &R    \cr
2019 03 01.406 &  8543.906 &  22.5  &0.4  &R   \cr
2019 03 02.282 &  8544.782 &  22.1  &0.4  &R    \cr
2019 03 05.405 &  8547.905 &  22.4  &0.4  &R   \cr
2019 03 06.405 &  8548.905 &  22.6  &0.4  &R   \cr
2019 03 08.403 &  8550.903 &  21.9  &0.3  &I   \cr
\cutinhead{2019-nova2 = 2019-03a}
2018 03 26.207 &  8203.707 & [24.0  &\dots&R    \cr
2019 02 06.394 &  8520.894 & [22.7  &\dots&R    \cr
2019 02 18.399 &  8532.899 &  22.0  &0.4  &R    \cr
2019 03 01.406 &  8543.906 &  22.2  &0.4  &R   \cr
2019 03 02.282 &  8544.782 &  22.0  &0.2  &R    \cr
2019 03 04.404 &  8546.904 &  21.9  &0.25 &R   \cr
2019 03 05.405 &  8547.905 &  22.2  &0.3  &R   \cr
2019 03 06.405 &  8548.905 &  22.0  &0.3  &R   \cr
2019 03 08.403 &  8550.903 &  21.9  &0.3  &I   \cr
2019 03 10.386 &  8552.886 &  22.1  &0.25 &R   \cr
2019 03 12.387 &  8554.887 &  21.64 &0.09 &R   \cr
2019 03 13.406 &  8555.906 &  21.65 &0.09 &R   \cr
2019 03 14.363 &  8556.863 &  21.56 &0.08 &R   \cr
2019 03 21.245 &  8563.745 &  21.9  &0.1  &R    \cr
\cutinhead{Probable SN in anonymous galaxy in the field of M83}
2012 06 01.138 &  6079.638 & [22.8  &\dots&R   \cr
2013 01 12.374 &  6304.874 & [22.5  &\dots&R   \cr
2013 01 17.368 &  6309.868 &  21.9  &0.25 &R   \cr
2013 01 18.360 &  6310.860 &  21.6  &0.2  &R   \cr
2013 01 18.380 &  6310.880 &  21.6  &0.3  &I   \cr
2013 01 19.387 &  6311.887 &  21.2  &0.3  &R   \cr
2013 01 22.373 &  6314.873 &  21.3  &0.15 &R   \cr
2013 01 23.387 &  6315.887 &  21.3  &0.15 &R   \cr
2013 01 26.382 &  6318.882 &  21.1  &0.15 &R   \cr
2013 01 28.389 &  6320.889 &  20.9  &0.2  &R   \cr
2013 01 29.378 &  6321.878 &  20.9  &0.15 &R   \cr
2013 01 29.392 &  6321.892 &  20.5  &0.15 &I   \cr
2013 01 31.362 &  6323.862 &  21.0  &0.1  &R    \cr
2013 02 01.308 &  6324.808 &  20.9  &0.15 &R    \cr
2013 02 03.327 &  6326.827 &  21.0  &0.1  &R    \cr
2013 02 05.390 &  6328.890 &  20.8  &0.25 &R   \cr
2013 02 07.397 &  6330.897 &  21.5  &0.3  &R   \cr
2013 02 08.395 &  6331.895 &  21.5  &0.3  &R   \cr
2013 02 09.381 &  6332.881 &  21.0  &0.35 &R    \cr
2013 02 11.387 &  6334.887 &  21.2  &0.15 &R    \cr
2013 02 12.389 &  6335.889 &  21.3  &0.1  &R   \cr
2013 02 13.397 &  6336.897 &  21.4  &0.15 &R   \cr
2013 02 14.398 &  6337.898 &  21.3  &0.15 &R   \cr
2013 02 16.396 &  6339.896 &  21.5  &0.2  &R   \cr
2013 02 18.388 &  6341.888 &  21.6  &0.2  &R   \cr
2013 02 19.385 &  6342.885 &  21.7  &0.2  &R    \cr
2013 02 20.404 &  6343.904 &  21.8  &0.3  &R   \cr
2013 02 22.362 &  6345.862 &  21.7  &0.1  &R    \cr
2013 02 24.349 &  6347.849 &  21.8  &0.15 &R    \cr
2013 02 25.386 &  6348.886 &  22.0  &0.15 &R    \cr
2013 03 01.338 &  6352.838 &  22.1  &0.15 &R    \cr
2013 03 02.288 &  6353.788 &  21.4  &0.15 &R    \cr
2013 03 04.397 &  6355.897 &  22.3  &0.2  &R   \cr
2013 03 06.401 &  6357.901 &  22.1  &0.2  &R   \cr
2013 03 09.404 &  6360.904 &  22.4  &0.2  &R   \cr
2013 03 14.244 &  6365.744 &  22.6  &0.25 &R   \cr
2013 03 24.377 &  6375.877 &  22.8  &0.3  &R   \cr
2013 04 06.249 &  6388.749 &  22.6  &0.4  &R    \cr
2013 04 07.301 &  6389.801 &  23.0  &0.4  &R    \cr
2013 04 08.360 &  6390.860 &  22.7  &0.35 &R   \cr
\enddata
\end{deluxetable}

\end{document}